\documentclass[12pt]{iopart}
\usepackage[british]{babel}
\usepackage{graphicx}  
\usepackage{dcolumn}
\usepackage{bm}         
\usepackage{verbatim}  
\usepackage{color}
\usepackage[dvipsnames]{xcolor}
\usepackage[perpage]{footmisc}
\usepackage{cite}

\usepackage{iopams}  

\expandafter\let\csname equation*\endcsname\relax
\expandafter\let\csname endequation*\endcsname\relax
\usepackage{physics} 

\usepackage{xcolor}	 
\definecolor{myblue}{RGB}{34,31,150}

\usepackage[breaklinks=true,colorlinks=true,linkcolor=myblue,urlcolor=myblue,citecolor=myblue]{hyperref}

\makeatletter
\newcommand*{\transpose}{%
  {\mathpalette\@transpose{}}%
}
\newcommand*{\@transpose}[2]{%
  \raisebox{\depth}{$\m@th#1\intercal$}%
}
\makeatother

\begin{document}

\title[Quantum sensing networks for the estimation of linear functions]{Quantum sensing networks for the estimation of linear functions}

\author{Jes\'{u}s Rubio$^{1}$, Paul A Knott$^2$, Timothy J Proctor$^3$ and Jacob A Dunningham$^4$}
\address{$^1$Department of Physics and Astronomy, University of Exeter, Stocker Road, Exeter EX4 4QL UK}
\address{$^2$Centre for the Mathematics and Theoretical Physics of Quantum Non-Equilibrium Systems (CQNE), School of Mathematical Sciences, University of Nottingham, University Park, Nottingham NG7 2RD UK}
\address{$^3$Sandia National Laboratories, Livermore CA 94550 USA}
\address{$^4$Department of Physics and Astronomy, University of Sussex, Brighton BN1 9QH UK}
\eads{\mailto{J.Rubio-Jimenez@exeter.ac.uk}, \mailto{J.Dunningham@sussex.ac.uk}}
\vspace{10pt}
\begin{indented}
\item[](Dated: \today)
\end{indented}

\begin{abstract}

The theoretical framework for networked quantum sensing has been developed to a great extent in the past few years, but there are still a number of open questions. Among these, a problem of great significance, both fundamentally and for constructing efficient sensing networks, is that of the role of inter-sensor correlations in the \emph{simultaneous} estimation of multiple linear functions, where the latter are taken over a collection local parameters and can thus be seen as global properties. In this work we provide a solution to this when each node is a qubit and the state of the network is sensor-symmetric. First we derive a general expression linking the amount of inter-sensor correlations and the geometry of the vectors associated with the functions, such that the asymptotic error is optimal. Using this we show that if the vectors are clustered around two special subspaces, then the optimum is achieved when the correlation strength approaches its extreme values, while there is a monotonic transition between such extremes for any other geometry. Furthermore, we demonstrate that entanglement can be detrimental for estimating non-trivial global properties, and that sometimes it is in fact irrelevant. Finally, we perform a non-asymptotic analysis of these results using a Bayesian approach, finding that the amount of correlations needed to enhance the precision crucially depends on the number of measurement data. Our results will serve as a basis to investigate how to harness correlations in networks of quantum sensors operating both in and out of the asymptotic regime.

\end{abstract}

%
%
%
%

\section{Introduction}
\label{intro}

\begin{figure}[t]
\centering
\includegraphics[trim={3.5cm 2.75cm 2.05cm 1.75cm},clip,width=6cm]{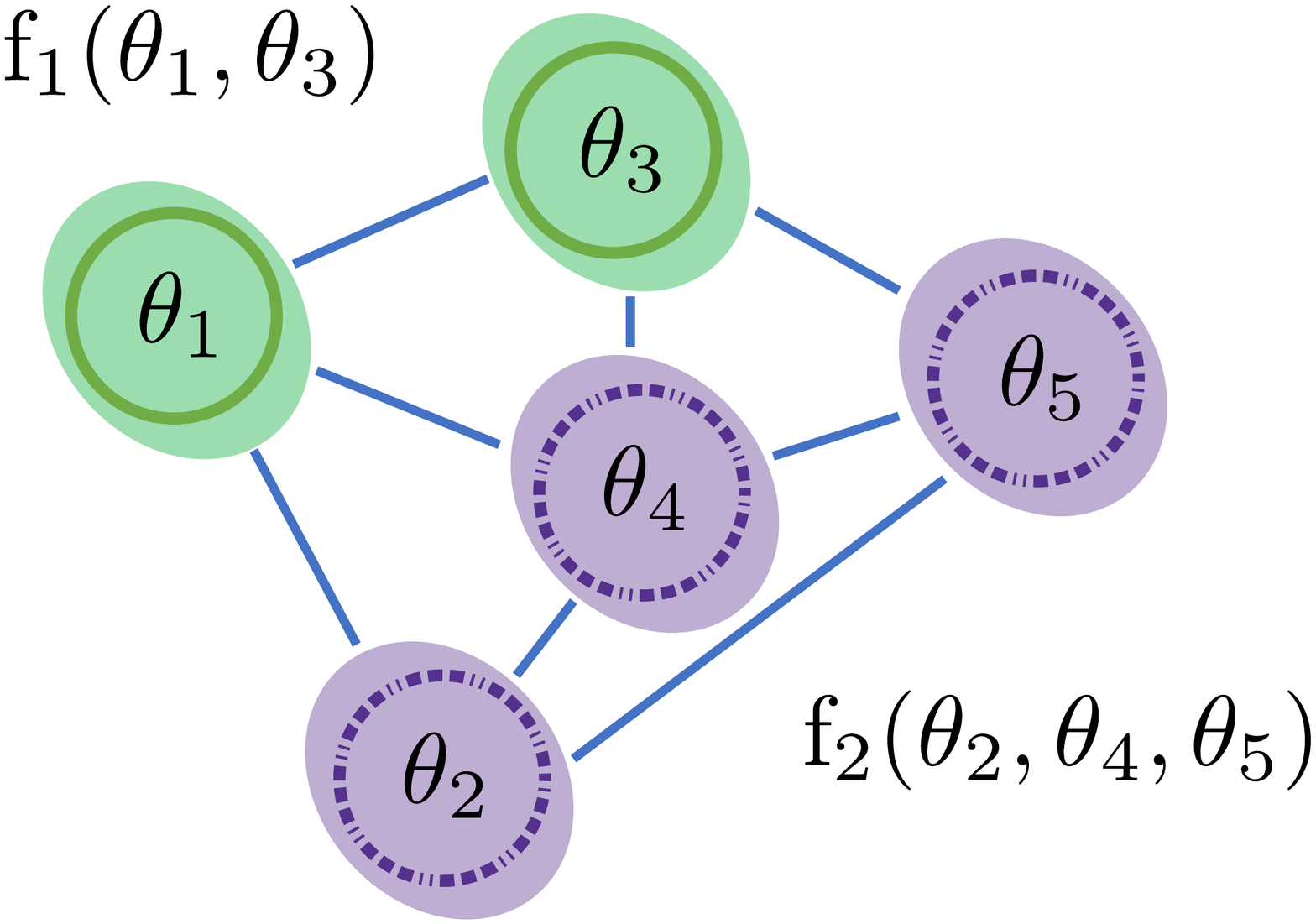}
	\caption{A network of $d = 5$ sensors. The parameters $\boldsymbol{\theta} = (\theta_1, \dots, \theta_5)$ represent local properties, since each of them is locally encoded in a single sensor. On the contrary, $f_1(\theta_1, \theta_3)$ and $f_2(\theta_2, \theta_4, \theta_5)$ are global properties associated with sensors $1$ and $3$ (green solid lines) and sensors $2$, $4$ and $5$ (purple dashed lines), respectively. }
\label{networkexample}
\end{figure}

An important task in quantum information science is to devise protocols for multi-parameter metrology and estimation by exploiting the quantum properties of light and matter. This problem has been widely explored not only in a theoretical fashion \cite{yuen1973, holevo1973b, holevo1973, helstrom1974, helstrom1976, paris2009, holevo2011, gill2011, zhang2014, berry2015, tsang2016, sammy2016compatibility, pezze2017simultaneous, hall2018, gessner2018, jesus2019b, yang2019, carollo2019, albarelli2019, tsang2019dec, sidhu2019dec, albarelli2019novA}, but also in applications \cite{chiara2003, spagnolo2012, chiribella2012, humphreys2013, vidrighin2014, zhang2014, baumgratz2016, knott2016local, gagatos2016gaussian, jasminder2016, proctor2017networked, proctor2017networkedshort, szczykulska2017, zhang_lu2017, zhuang2017, altenburg2018, jasminder2018, gessner2018, jesus2019b, albarelli2019novB} and experiments \cite{vidrighin2014, roccia2018, polino2018,  valeri2020}. As a result, new practical ways of enhancing our estimation schemes have recently emerged \cite{Szczykulska2016, liu2019, jesus2019thesis, jasminder2019, rafal2020, polino2020}. These protocols are normally formulated on the basis of $d$ unknown parameters $\boldsymbol{\theta} = (\theta_1, \dots, \theta_d)$ that arise naturally in the description of the system at hand, and in many cases these are the quantities of interest. However, sometimes we may wish or need to find $l$ new quantities that are functions of $\boldsymbol{\theta}$, that is, $\boldsymbol{f}(\boldsymbol{\theta}) = (f_1(\boldsymbol{\theta}), \dots, f_l(\boldsymbol{\theta}))$. This is the case, in particular, when we analyse global properties in a quantum sensing network \cite{proctor2017networked, proctor2017networkedshort}, which is a model for spatially distributed sensing \cite{jasminder2019} and the main focus of this work. Indeed, in \cite{proctor2017networked, proctor2017networkedshort} this model is defined as an array of quantum sensors where one or several parameters are locally encoded in each of them, and while a property of the network is said to be \emph{local} if it is represented by parameters at a single sensor, a \emph{global} property is thought of as a non-trivial function of two or more parameters at different sensors. Here we consider that a single parameter $\theta_i$ is encoded in the $i$-th sensor, so that $\boldsymbol{\theta}$ is a collection of local properties, and we assume that both parameters and functions are real-valued quantities. See figure \ref{networkexample} for a schematic representation.

Networked scenarios where global properties are relevant provide a natural testbed to identify the potential usefulness of entanglement in a broad range of multi-parameter schemes \cite{proctor2017networked, altenburg2018}. Within this context, the optimal estimation of a single function $f(\boldsymbol{\theta})$ has been extensively studied \cite{proctor2017networked, proctor2017networkedshort, eldredge2018, altenburg2018, ge2018, zhuang2018, qian2019,  zachary2019thesis, sekatski2019, gatto2019, guo2019, oh2019, zhuang2019, jasminder2019}, and it has been established that one can find entangled states that beat the best separable probe when that function is linear \cite{proctor2017networked, proctor2017networkedshort}. In addition, Eldredge \emph{et al.} \cite{eldredge2018} derived a bound on the error for this scenario that was later generalised to accommodate a single analytical function \cite{qian2019}, which can also be estimated with an enhanced precision when there is entanglement, while Gross and Caves \cite{gross2020} have reexamined the linear case using an elegant geometric approach. On the opposite extreme, it has been shown that a collection of $l = d$ linear functions that generates an orthogonal transformation (i.e., $\boldsymbol{f}(\boldsymbol{\theta}) = V^{\transpose} \boldsymbol{\theta}$ with $V V^{-1}=\mathbb{I}$) can be estimated optimally with a local strategy \cite{proctor2017networked, altenburg2018}.           

Beyond these two types of global properties, the simultaneous estimation of $l > 1$ linear but otherwise arbitrary real functions has been a less travelled path. There exist generic bounds for this problem (see, e.g, \cite{proctor2017networked, li2019}), which in practice may arise in scenarios such as the estimation of phase differences \cite{knott2016local, li2019}. However, how quantum correlations may help for linear functions with arbitrary geometry has not been examined in detail. Given that this represents a richer regime than the $l = 1$ and $l = d$ with orthogonal functions cases, it can be argued that answering this question is essential for further progress in networked quantum metrology. 

While a general answer is beyond the scope of our methods, here we obtain a definite solution for a subclass of schemes with sensor-symmetric pure qubit states, which we introduce in section \ref{ournetwork}. Using the Helstrom Cram\'{e}r-Rao bound and the associated quantum Fisher information matrix, in section \ref{asymptoticresults} we derive a general expression linking the geometry of the vector components associated with the functions and the strength of the inter-sensor correlations, such that the uncertainty in the asymptotic regime of many trials is optimal. Moreover, we show that there exists a physical state for many of the optimal configurations that our formula predicts. Equipped with this, we then derive a number of important results. First we find that the largest amounts of correlations are associated, for sensor-symmetric states, with two special subspaces: the direction of the vector of ones $\boldsymbol{1}^\transpose \equiv (1, 1, \dots)$, and the subspace orthogonal to it. This connection between entanglement in a pure state and how much the vectors are clustered around certain directions was precisely one of the open questions identified in \cite{proctor2017networked}, and our findings contribute towards its solution. In addition, we demonstrate that entanglement can be detrimental for estimating global properties other than those associated with orthogonal transformations, while a three-sensor network reveals that entanglement is sometimes irrelevant. This is consistent with the fact that the asymptotic uncertainty only depends on correlations of a pairwise nature, and thus other forms of entanglement do not affect the asymptotic error.

On the other hand, it is known that strategies with a good asymptotic precision found by optimising the Cram\'{e}r-Rao bound sometimes have a particularly poor performance when the number of trials is very low (see, e.g, \cite{jesus2017}). In fact, there is compelling evidence of the existence of a potential trade-off between the performances in the asymptotic and non-asymptotic regimes \cite{jesus2018}. In view of this, a non-asymptotic analysis of our findings for sensing networks is in order. To do it, in section \ref{hybrid} we propose a multi-parameter Bayesian procedure that generalises its single-parameter counterpart in \cite{jesus2017}, and in section \ref{bayesianresults} we utilise it to examine the non-asymptotic properties of some of our results in section \ref{asymptoticresults}. Our central insight here is that trading a part of the asymptotic enhancement is sometimes associated with an improved performance in the non-asymptotic regime \emph{also} in networked quantum metrology, and in general we find that the amount of correlations needed to enhance the precision crucially depends on the amount of data that has been collected. Due to the more complex (and often numerical) nature of Bayesian calculations, this study is restricted to the $d=2$ case, although in section \ref{conclusions} we discuss some potential directions to overcome this limitation. To the best of our knowledge, this work, together with \cite{sekatski2019, jesus2019b}, constitutes one of the first Bayesian studies of a network of quantum sensors in this context.

Our approach to the simultaneous estimation of linear functions in a scheme for distributed quantum sensing will serve as a basis to investigate how to harness correlations in multi-parameter schemes, operating both in and out of the asymptotic regime. Since the construction of entangled networks is likely to be difficult in practice, these insights may prove to be crucial in the study and implementation of quantum sensing networks that operate with a realistic amount of data. 
\section{Formulation of the problem}

\subsection{Physical scheme and available information}\label{ournetwork}

Consider a network of $d$ qubit sensors prepared in some initial state $\rho_0 = \ketbra{\psi_0}$, with
\begin{equation}
\ket{\psi_0} = \sum_{i_1 \dots i_d=0}^1 a_{i_1 \dots i_d} \ket{i_0, i_1 \dots i_d},
\label{genericprobe}
\end{equation}
$\sum_{i_1 \dots i_d=0}^1 |a_{i_1 \dots i_d}|^2 = 1$, and the basis elements $\bra{0}_j = (1, 0)$ and $\bra{1}_j = (0, 1)$ for the $j$-th sensor. In addition, suppose we encode $d$ local parameters $\boldsymbol{\theta} = (\theta_1, \dots, \theta_d)$, one per sensor, as $\rho(\boldsymbol{\theta})=\mathrm{e}^{-i \boldsymbol{K}\cdot\boldsymbol{\theta}}\rho_0\hspace{0.1em} \mathrm{e}^{i \boldsymbol{K}\cdot\boldsymbol{\theta}}$, where $\boldsymbol{K} = (K_1, \dots, K_d)$, each generator $K_i$ has the form
\begin{equation}
2 K_i = \mathbb{I}_1\otimes\cdots\otimes\mathbb{I}_{i-1}\otimes \sigma_z\otimes\mathbb{I}_{i+1}\otimes\cdots\otimes\mathbb{I}_d \equiv \sigma_{z, i},
\end{equation}
and 
\begin{equation}
\sigma_z =
\begin{pmatrix}
1 & 0 \\
0 & -1
\end{pmatrix},\hspace{0.4em}
\mathbb{I}
_i = 
\begin{pmatrix}
1 & 0 \\
0 & 1
\end{pmatrix}.
\end{equation}
This is an instance of the type of unitary encoding that arises in spatially distributed sensing \cite{proctor2017networkedshort, proctor2017networked}, and while it is separable, i.e., 
\begin{equation}
\mathrm{exp}\left(-i \boldsymbol{K}\cdot\boldsymbol{\theta}\right) = \mathrm{e}^{-i \sigma_z \theta_1/2}\otimes \cdots \otimes \mathrm{e}^{-i \sigma_z \theta_d/2},
\label{networksunitary}
\end{equation}
in principle we allow for entangled pure states and any general measurement acting on all the sensors at once. When the state and the measurement present no quantum correlations, we say that the scheme implements a \emph{local strategy}. Otherwise we have a \emph{global strategy}. We also note that
\begin{equation}
[K_i, K_j] = [\sigma_{z, i},\sigma_{z, j}]/4 = 0,
\label{commcondition}
\end{equation}
which is a useful feature of this system because it will allow us to saturate the asymptotic bound in section \ref{hybrid}.

To introduce the subclass of \emph{sensor-symmetric states} that we will exploit, first we recall that the strength of correlations between any pair of sensors, which we call \emph{inter-sensor correlations}, may be quantified as \cite{knott2016local, proctor2017networked} 
\begin{equation}
\mathcal{J}_{ij} = \frac{\langle K_i K_j \rangle - \langle K_i \rangle\langle K_j \rangle}{\Delta K_i \Delta K_j},
\label{multicorrelations}
\end{equation}
for $i \neq j$, where $\Delta K_i^2 = \langle K_i^2 \rangle - \langle K_i \rangle^2$ and we use the notation
$\langle \Box \rangle \equiv \bra{\psi_0} \Box \ket{\psi_0}$. Furthermore, $\mathcal{J}_{ij}$ in equation (\ref{multicorrelations}) is bounded as $-1 \leqslant \mathcal{J}_{ij} \leqslant 1$. Using this quantifier, we define sensor-symmetric states as those satisfying
\begin{align}
v = \langle K_i^2 \rangle - \langle K_i \rangle^2, ~~ c =\langle K_i K_j \rangle - \langle K_i \rangle\langle K_j \rangle
\label{sensorsymcon}
\end{align}
for all $i$, $j$, where $c$ and $v$ are fixed values that characterise the preparation of the network and the encoding of the parameters. In turn, equation (\ref{multicorrelations}) becomes $\mathcal{J}_{ij}=\mathcal{J} = c/v$, also for all $i\neq j$, and for our qubit model we see that
\begin{align}
4v = \langle \sigma_{z, i}^2 \rangle - \langle \sigma_{z, i} \rangle^2 = 1 - \langle \sigma_{z, i} \rangle^2,~~
4c = \langle \sigma_{z, i} \sigma_{z, j} \rangle - \langle \sigma_{z, i} \rangle\langle \sigma_{z, j} \rangle,
\label{sensorsymconqubit}
\end{align}
where $0 \leqslant 4v \leqslant 1$ due to the fact that the eigenvalues of $\sigma_z$ are $\pm 1$ and thus $\abs{\langle \sigma_{z, i} \rangle} \leqslant 1$. This definition in terms of the conditions in equation (\ref{sensorsymcon}) is a way of generalising the notion of path-symmetric states in optical interferometry \cite{knott2016local, HofmannHolger2009, sahota2015}, and it motivates our choice of initial probe.

The final piece required before we can formulate the estimation problem of interest is to establish what prior information is available. The properties of the network that we wish to estimate are those that can be modelled linearly as 
\begin{equation}
\boldsymbol{f}(\boldsymbol{\theta}) = (f_1(\boldsymbol{\theta}), \dots, f_l(\boldsymbol{\theta})) = V^{\transpose} \boldsymbol{\theta} + \boldsymbol{a},
\end{equation}
where $V$ is a $(d\times l)$ matrix and $\boldsymbol{a}$ is a column vector with $l$ components. We consider that the form of these functions is known and so there is no uncertainty associated with the matrix $V$ or the vector $\boldsymbol{a}$. Furthermore, we assume that the unknown parameters $\boldsymbol{\theta}$ can be initially thought of as independent in the statistical sense, such that there are no prior correlations between them, and we suppose that the magnitude of the $i$-th parameter can be found somewhere within an interval of width $W_{0,i}$ centred around $\bar{\theta}_i$, which is a moderate amount of prior knowledge \cite{jesus2018, jesus2019thesis, demkowicz2011}. This state of information can be represented by the separable prior probability
\begin{equation}
p(\boldsymbol{\theta})= 1/\left(\prod_{i=1}^d W_{0,i}\right), 
\label{multiprior}
\end{equation}
for $\boldsymbol{\theta}\in [\bar{\theta}_1 - W_{0, 1}/2, \bar{\theta}_1 + W_{0, 1}/2]\times \cdots \times [\bar{\theta}_d - W_{0, d}/2, \bar{\theta}_d + W_{0,d}/2]$, and zero otherwise. Equivalently, equation (\ref{multiprior}) may also be written as $p(\boldsymbol{\theta}) = 1/\Delta_0$, with hypervolume $\Delta_0 = \prod_{i=1}^d W_{0,i}$ centred around $\boldsymbol{\bar{\theta}} = (\bar{\theta}_1,\dots,\bar{\theta}_d)$. The interested reader will find in \ref{multipriorappsec} a way of justifying this prior from the perspective of the so-called \emph{objective} version of the Bayesian framework.

\subsection{Estimation method: a hybrid approach}\label{hybrid}

Starting with the transformed network state $\rho(\boldsymbol{\theta})$ in section \ref{ournetwork}, the next step is to consider $\mu$ identical and independent measurements on this system, which we see as \emph{trials} or \emph{repetitions}. In particular, the $i$-th measurement is represented by a POVM $E(m_i)$ with outcome $m_i$, and the probability of this process generating the outcomes $\boldsymbol{m} = (m_1, \dots, m_\mu)$ is given by the likelihood function
\begin{equation}
p(\boldsymbol{m}|\boldsymbol{\theta}) = \prod_{i=1}^\mu p(m_i|\boldsymbol{\theta}) = \prod_{i=1}^\mu \mathrm{Tr}\left[E(m_i) \rho(\boldsymbol{\theta}) \right].
\end{equation} 
Since the form of the functions $\boldsymbol{f}(\cdot)$ has been assumed to be known, it is appropriate to construct their estimators as 
\begin{equation}
\boldsymbol{\tilde{f}}(\boldsymbol{m}) = \boldsymbol{f}[\boldsymbol{\tilde{\theta}}(\boldsymbol{m})] = (f_1[\boldsymbol{\tilde{\theta}}(\boldsymbol{m})], \dots, f_l[\boldsymbol{\tilde{\theta}}(\boldsymbol{m})]) = V^\transpose \boldsymbol{\tilde{\theta}}(\boldsymbol{m}) + \boldsymbol{a},
\label{funest}
\end{equation}
where $\boldsymbol{\tilde{\theta}}(\boldsymbol{m}) = (\tilde{\theta}_1(\boldsymbol{m}), \dots, \tilde{\theta}_d(\boldsymbol{m}))$ are the estimators for the parameters $\boldsymbol{\theta}$, and we evaluate the uncertainty of our estimates $\boldsymbol{\tilde{f}}(\boldsymbol{m})$ as
\begin{align}
\bar{\epsilon}_{\mathrm{mse}} &= \int d\boldsymbol{\theta} d\boldsymbol{m} ~p(\boldsymbol{\theta})p(\boldsymbol{m}|\boldsymbol{\theta})~\mathrm{Tr} \lbrace \mathcal{W} [\boldsymbol{\tilde{f}}(\boldsymbol{m}) - \boldsymbol{f}(\boldsymbol{\theta}) ] [\boldsymbol{\tilde{f}}(\boldsymbol{m}) - \boldsymbol{f}(\boldsymbol{\theta})]^\transpose \rbrace,
\label{msegenfunctions}
\end{align}
where $p(\boldsymbol{\theta})$ is the prior, $\mathcal{W}= \mathrm{diag}(w_1, \cdots, w_l)$ is a weighting matrix, $w_i \geqslant 0$ represents the relative importance of estimating the $i$-th parameter, and $\mathrm{Tr}(\mathcal{W}) = 1$. 
Importantly, although a square error is generally not suitable for quantities associated with topologies other than that for the real line, it can still be a good approximation to the uncertainty for other topologies when the prior knowledge about $\boldsymbol{\theta}$ is moderate or high (see, e.g, \cite{rafal2015, jesus2017, friis2017, jesus2018, jesus2019thesis, rafal2020}), which is our case. 

By using equations (\ref{multiprior} - \ref{funest}) and the network configuration in section \ref{ournetwork}, equation (\ref{msegenfunctions}) becomes
\begin{align}
\bar{\epsilon}_{\mathrm{mse}} = &\int \frac{d\boldsymbol{\theta} d\boldsymbol{m}}{\Delta_0}~ \prod_{i=1}^\mu \mathrm{Tr}\left[E(m_i)\hspace{0.2em} \mathrm{e}^{-i \boldsymbol{K}\cdot\boldsymbol{\theta}}\rho_0\hspace{0.1em} \mathrm{e}^{i \boldsymbol{K}\cdot\boldsymbol{\theta}} \right]
\nonumber \\
&\times \mathrm{Tr}\lbrace \mathcal{W} V^\transpose [\boldsymbol{\tilde{\theta}}(\boldsymbol{m}) - \boldsymbol{\theta}]  [\boldsymbol{\tilde{\theta}}(\boldsymbol{m}) - \boldsymbol{\theta}]^\transpose V \rbrace
\label{msefunctions}
\end{align}
for our system. We note that this error does not depend on $\boldsymbol{a}$, so that we can set $\boldsymbol{a} = \boldsymbol{0}$ without loss of generality. Hence, from now on the functions are $\boldsymbol{f}(\boldsymbol{\theta}) = V^{\transpose}\boldsymbol{\theta}$ and the coefficients are encoded in the columns of $V$. 

Ideally, we would like to minimise the error in equation (\ref{msefunctions}) with respect to the estimators $\boldsymbol{\tilde{\theta}}(\boldsymbol{m})$, the measurement scheme $E(m_i)$ and the initial sensor-symmetric state $\rho_0$, so that we can find the optimal configuration of the network and study its properties. Since, in general, this is a very challenging problem, in this work we follow an approximate procedure that combines asymptotic and non-asymptotic optimisations. We now describe this \emph{hybrid} approach and how to use it for our analysis of sensing networks (a discussion of other methods in the literature can be found in \ref{multioptimisation}). 

On the one hand, equation (\ref{msefunctions}) can be minimised with respect to $\boldsymbol{\tilde{\theta}}(\boldsymbol{m})$ in a straightforward way (e.g., using calculus of variations; see \cite{jesus2019b, jaynes2003}). This provides the familiar result that 
\begin{equation}
\boldsymbol{\tilde{\theta}}(\boldsymbol{m}) = \int d\boldsymbol{\theta} ~p(\boldsymbol{\theta}|\boldsymbol{m})~ \boldsymbol{\theta}
\label{optestloc}
\end{equation}
are the optimal estimators \cite{kay1993, jaynes2003}, where $p(\boldsymbol{\theta}|\boldsymbol{m}) = p(\boldsymbol{m}|\boldsymbol{\theta})/[\Delta_0 \hspace{0.2em} p(\boldsymbol{m})]$ is the posterior probability and $p(\boldsymbol{m}) = \int d\boldsymbol{\theta}\hspace{0.2em} p(\boldsymbol{m}|\boldsymbol{\theta})/\Delta_0$. As a consequence, inserting equation (\ref{optestloc}) in equation (\ref{msefunctions}) we have that
\begin{equation}
\bar{\epsilon}_{\mathrm{mse}} \geqslant \sum_{i=1}^l w_i \int d\boldsymbol{m}~ p(\boldsymbol{m}) \left\lbrace \int d\boldsymbol{\theta} p(\boldsymbol{\theta}|\boldsymbol{m}) f_i^2(\boldsymbol{\theta}) - \left[\int d\boldsymbol{\theta} p(\boldsymbol{\theta}|\boldsymbol{m}) f_i(\boldsymbol{\theta})\right]^2 \right\rbrace \equiv \epsilon_{\mathrm{opt}}^c,
\label{msefunctionsmin}
\end{equation}
where $f_i(\boldsymbol{\theta}) = \sum_{j=1}^d V_{ji} \theta_j$. This is the optimal uncertainty based on the probabilities that emerge from the measurements in a given quantum strategy ($E(m_i)$ plus $\rho_0$), and is valid and exact for any number of trials $\mu$.

On the other hand, we may select the quantum strategy such that it is optimal in the asymptotic regime of many trials, where $\mu \gg 1$. First we recall that, if the true values $\boldsymbol{\theta}'$ lie within the prior hypervolume $\Delta_0$, and the likelihood $p(\boldsymbol{m}|\boldsymbol{\theta})$, which we assume to be sufficiently regular, becomes concentrated around $\boldsymbol{\theta}'$ as $\mu$ grows, then the posterior probabiliy $p(\boldsymbol{\theta}|\boldsymbol{m})$ can be approximated as a multivariate Gaussian density, and the uncertainty $\epsilon_{\mathrm{opt}}^c$ in equation (\ref{msefunctionsmin}) satisfies \cite{jaynes2003, cox2000, bernardo1994}
\begin{align}
\epsilon_{\mathrm{opt}}^c \approx \int \frac{d\boldsymbol{\theta}'}{\mu \Delta_0} \hspace{0.15em}\mathrm{Tr}\left[\mathcal{W} V^\transpose F(\boldsymbol{\theta}')^{-1} V \right] \equiv \epsilon_{\mathrm{asym}}^c,
\label{ccrb}
\end{align}
where 
\begin{align}
F(\boldsymbol{\theta}) = \int \frac{dm}{p(m|\boldsymbol{\theta})} \left[\frac{\partial p(m|\boldsymbol{\theta})}{\partial \boldsymbol{\theta}} \right] \left[\frac{\partial p(m|\boldsymbol{\theta})}{\partial \boldsymbol{\theta}} \right]^\transpose
\label{fim}
\end{align}
is the Fisher information matrix for a single trial with outcome $m$ (for a derivation of this approximation, see, e.g., \cite{jaynes2003, cox2000, bernardo1994} and section 6.2.2 of \cite{jesus2019thesis}, and \cite{lecam1986, vaart1998, gill2011} for a rigorous treatment). At the same time, given that the form of the unitary encoding is $\mathrm{exp}(-i \boldsymbol{K}\cdot\boldsymbol{\theta})$ and the state $\rho_0 = \ketbra{\psi_0}$ is pure, the Helstrom Cram\'{e}r-Rao bound establishes the matrix inequality \cite{Szczykulska2016, liu2019, jasminder2019, rafal2020}
\begin{equation}
F(\boldsymbol{\theta})^{-1} \geqslant F_q^{-1}, ~\text{with}~(F_q)_{ij} =  4\left( \langle  \psi_0 | K_i K_j |  \psi_0 \rangle - \langle  \psi_0 | K_i |  \psi_0 \rangle \langle  \psi_0 | K_j |  \psi_0 \rangle \right),
\label{helstrombound}
\end{equation}
$F_q$ being the quantum counterpart of the information matrix. Then, the combination of equations (\ref{msefunctionsmin}), (\ref{ccrb}) and (\ref{helstrombound}) implies that, in the asymptotic regime,
\begin{align}
\bar{\epsilon}_\mathrm{mse} \geqslant \epsilon_{\mathrm{opt}}^c \approx \epsilon_{\mathrm{asym}}^c \geqslant \frac{1}{\mu} \mathrm{Tr}\left(\mathcal{W} V^\transpose F_q^{-1} V \right) \equiv \bar{\epsilon}_{\mathrm{cr}}.
\label{qcrb}
\end{align}

The quantum Cram\'{e}r-Rao bound $\bar{\epsilon}_{\mathrm{cr}}$ in equation (\ref{qcrb}) is a function of $\rho_0$ only, since $\boldsymbol{K}$, $V$, $\mathcal{W}$ and $\mu$ are fixed, and it does not depend on the measurement. As such, if we choose the POVM $E(m_i)$ for the $i$-th repetition such that $\epsilon_{\mathrm{asym}}^c = \bar{\epsilon}_{\mathrm{cr}}$, then that measurement will be asymptotically optimal. It can be shown that a measurement such that $F(\boldsymbol{\theta}) = F_q$ (and thus $\epsilon_{\mathrm{asym}}^c = \bar{\epsilon}_{\mathrm{cr}}$) always exists when the generators $\boldsymbol{K}$ commute with each other \cite{sammy2016compatibility, pezze2017simultaneous}, and equation (\ref{commcondition}) demonstrates that this is indeed satisfied by our qubit network. Hence, we will use this criterion to construct the POVM. Regarding the optimisation of the state, we will proceed by first calculating $\bar{\epsilon}_{\mathrm{cr}}$ as a function of the properties that characterise the sensor-symmetric state $\rho_0$, which, as we will see, are the variance $v$ and the correlation strength $\mathcal{J}$, and then minimising the resulting bound with respect to the pair $(v, \mathcal{J})$. Once we know the optimal estimators
\begin{equation}
\boldsymbol{\tilde{f}}(\boldsymbol{m}) = V^\transpose \int d\boldsymbol{\theta} ~p(\boldsymbol{\theta}|\boldsymbol{m})~ \boldsymbol{\theta}
\label{fundoptest}
\end{equation}
and the asymptotically optimal state and measurement as prescribed above, we can complete the estimation by inserting these in the Bayesian uncertainty for $\mu$ repetitions in equation (\ref{msefunctions}), which here will be calculated numerically with the algorithm in section 6.2.3 of \cite{jesus2019thesis} (the reader interested in reproducing our numerical results will find the associated MATLAB code in Appendix C of the same work).

It is important to realise that our approach can fail when the asymptotic approximation is not valid. This could happen, for example, if the prior information provided within the hypervolume $\Delta_0$ is not sufficient to distinguish a single point \cite{jaynes2003, jesus2017}, or if the Fisher information matrix (classical or quantum) is singular. Therefore, we will concern ourselves with schemes where the information matrix is invertible, and, once we have found the asymptotically optimal quantum strategy, we will also check that the likelihood $p(\boldsymbol{m}|\boldsymbol{\theta})$ associated with it does not present ambiguities in the relevant portion of the parameter space. Nevertheless, note that, in general, a potentially ambiguous likelihood function or a singular $F(\boldsymbol{\theta})$ do not introduce any fundamental difficulty for Bayesian estimation itself (this will be demonstrated in section \ref{bayesianresults} with an example).

In summary, the estimation method that emerges from the previous discussion requires that we:
\begin{enumerate}
\item calculate the quantum Cram\'{e}r-Rao bound $\bar{\epsilon}_{\mathrm{cr}}$ and find the sensor-symmetric state that makes it minimal,
\item search for a POVM such that $\epsilon_{\mathrm{asym}}^c = \bar{\epsilon}_{\mathrm{cr}}$,
\item verify that the quantum strategy (state plus POVM) allows for unambiguous estimation given the prior information represented in equation (\ref{multiprior}),
\item calculate the optimal estimators for the linear functions in equation (\ref{fundoptest}), and
\item calculate the $\mu$-trial Bayesian uncertainty in equation (\ref{msefunctions}).
\end{enumerate}
While the protocols constructed in this way may not be optimal for low $\mu$, \cite{jesus2017} demonstrated that this technique can provide important information about the non-asymptotic regime in optical interferometry, and here we will show that this is also true for networked quantum sensing. Moreover, a very useful feature of our approach is that the analysis of the role of inter-sensor correlations emerging from (i, ii) will be relevant for researchers interested only in the Cram\'{e}r-Rao bound, while those that also require an analysis based on a finite number of repetitions will benefit from the insights arising from (iii - v). The next section is dedicated to the former.

\section{Asymptotic estimation of global properties}
\label{asymptoticresults}

\subsection{Estimation of arbitrary linear functions}
\label{sec:networksasym}

Our first step is to examine the quantum strategies that are optimal in the regime where the square error $\bar{\epsilon}_{\mathrm{mse}}$ converges to the quantum Cram\'{e}r-Rao bound $\bar{\epsilon}_{\mathrm{cr}} = \mathrm{Tr}(\mathcal{W} V^\transpose F_q^{-1} V)/\mu$ as $\mu$ grows. If we denote by $\lbrace \boldsymbol{e}_i \rbrace$ the basis components of the real space where $\mathcal{W}$, $V$ and $F_q$ are defined, with $\boldsymbol{e}_i^\transpose\boldsymbol{e}_j = \delta_{ij}$, then from equations (\ref{sensorsymconqubit}) and (\ref{helstrombound}) we have that
\begin{align}
F_q &= \sum_{i,j = 1}^d\left(\langle \sigma_{z, i} \sigma_{z, j} \rangle - \langle \sigma_{z, i}\rangle \langle \sigma_{z, j}\rangle \right) \boldsymbol{e}_i\boldsymbol{e}_j^\transpose = 4 \left(v \sum_{i=1}^d\boldsymbol{e}_i\boldsymbol{e}_i^\transpose + c \sum_{\underset{i\neq j}{i,j=1}}^d\boldsymbol{e}_i\boldsymbol{e}_j^\transpose  \right)
\nonumber \\
&= 4\left[(v-c)\mathbb{I} + c \mathcal{I}\right] = 4v\left[(1-\mathcal{J})\mathbb{I} + \mathcal{J} \mathcal{I}\right],
\label{fimsym}
\end{align}
where $\mathcal{I}$ is a $(d\times d)$ matrix of ones and $\mathbb{I}$ the $(d\times d)$ identity matrix. This is the quantum Fisher information matrix for sensor-symmetric states.

To invert $F_q$, we need to impose the condition of positive definiteness, which is equivalent to requiring that its eigenvalues are strictly positive. Expressing $\mathcal{I}$ as $\mathcal{I} = \boldsymbol{1}\boldsymbol{1}^\transpose$, where we recall that $\boldsymbol{1}^\transpose = (1, 1, \dots)$ is the vector of ones, the information matrix becomes $F_q = 4v\left[(1-\mathcal{J})\mathbb{I} + \mathcal{J} \boldsymbol{1}\boldsymbol{1}^\transpose\right]$. In that case, the characteristic equation for the eigenvalues $\lbrace \lambda \rbrace$ is
\begin{equation}
\mathrm{det} \left\lbrace 4v\left[\left(1-\mathcal{J}-\frac{\lambda}{4v}\right)\mathbb{I} + \mathcal{J}\boldsymbol{1}\boldsymbol{1}^\transpose\right] \right\rbrace = 0,
\label{characteristicfim}
\end{equation}
which upon using the identity $\mathrm{det}(X + \boldsymbol{y}\boldsymbol{z}^\transpose) = (1 + \boldsymbol{z}^\transpose X^{-1}\boldsymbol{y})\hspace{0.15em}\mathrm{det}(X)$, with $X = [4v(1-\mathcal{J}) - \lambda]\mathbb{I}$, $\boldsymbol{y} = 4v\mathcal{J}\boldsymbol{1}$ and $\boldsymbol{z}=\boldsymbol{1}$, implies that
\begin{equation}
\left\lbrace 4v\left[1+(d-1)\mathcal{J}\right]-\lambda\right\rbrace \left[4v\left(1-\mathcal{J}\right) - \lambda \right]^{d-1} = 0. 
\end{equation}
As a result, the eigenvalues of $F_q$ are $\lambda_1 = 4v[1+(d-1)\mathcal{J}]$, with multiplicity $1$, and $\lambda_2 = 4v(1-\mathcal{J})$, with multiplicity $d-1$, and by imposing that they are positive we conclude that $F_q$ is invertible when $1/(1-d)<\mathcal{J} < 1$. The rest of our calculations assume that $\mathcal{J}$ lies in such open interval under this assumption.

We can now calculate the inverse of $F_q$ in equation (\ref{fimsym}), which is \cite{proctor2017networked}
\begin{equation}
F_q^{-1} = \frac{\left[1 + (d-1)\mathcal{J}\right]\mathbb{I} - \mathcal{J}\mathcal{I}}{4v(1 -\mathcal{J})\left[1+(d-1)\mathcal{J}\right]}.
\label{fimsyminv}
\end{equation}
Utilising this result we find that the asymptotic uncertainty for the estimation of linear functions is given by
\begin{equation}
\bar{\epsilon}_{\mathrm{cr}} = \frac{\left[1 + (d-2)\mathcal{J}\right]\mathrm{Tr}\left(\mathcal{W} V^\transpose V \right) - \mathcal{J}\mathrm{Tr}\left(\mathcal{W} V^\transpose \mathcal{X} V \right)}{4\mu v (1 -\mathcal{J})[1+(d-1)\mathcal{J}]},
\label{symmetricfunctionsfirst}
\end{equation}
where we have introduced the $(d\times d)$ matrix $\mathcal{X} \equiv \mathcal{I} - \mathbb{I}$ to separate the contribution to the uncertainty due to the diagonal elements of $F_q^{-1}$, which are the errors for each of the parameters, from that of the rest of the matrix.

The expression in equation (\ref{symmetricfunctionsfirst}) shows that the uncertainty depends on three types of quantities: i) the number of repetitions $\mu$ and the number of parameters $d$, (ii) the combined properties of state and generators through the correlation strength $\mathcal{J}$ and the variance $v$, and (iii) two quantities, $\mathrm{Tr}\left(\mathcal{W} V^\transpose V \right)$ and $\mathrm{Tr}\left(\mathcal{W} V^\transpose \mathcal{X} V \right)$, that are defined in terms of the functions encoded in $V$ and the weighting matrix $\mathcal{W}$. The next step is to investigate the physical meaning of these two quantities in (iii).

By relabelling the vector formed by the components of the $j$-th linear function as $\boldsymbol{f}_j$ (i.e., $f_j(\boldsymbol{\theta}) = \sum_{i=1}^d V_{ij} \theta_i \equiv \boldsymbol{f}_j^\transpose\boldsymbol{\theta}$), we can rewrite the first quantity in a more suggestive form as
\begin{align}
\mathrm{Tr}\left(\mathcal{W} V^T V \right) &= \sum_{i, j = 1}^l \sum_{k=1}^d \left(\mathcal{W}\right)_{ij} V_{kj} V_{ki} = \sum_{j = 1}^l w_j \sum_{k=1}^d V_{kj} V_{kj} 
\nonumber \\
&= \sum_{j=1}^l w_j \boldsymbol{f}_j^\transpose \boldsymbol{f}_j = \sum_{j=1}^l w_j \abs{\boldsymbol{f}_j}^2.
\end{align}
where the norm in the last term is defined as $\abs{\boldsymbol{v}}^2 = \sum_k v_k^2$ for a real vector $\boldsymbol{v}$. This is the weighted sum of the squared magnitudes of the vectors associated with the linear functions. Since $V \mathcal{W} V^T$ is positive semi-definitive, and excluding the degenerate case where all the coefficients vanish, we have that $\mathrm{Tr}(\mathcal{W} V^T V) = \mathrm{Tr}(V \mathcal{W} V^T) > 0$. In addition, when the functions are normalised, that is, $|\boldsymbol{f}_i| = 1$ for $1 \leqslant i \leqslant l$, and recalling that $\mathrm{Tr}(\mathcal{W}) = \sum_{i=1}^l w_i= 1$, we have that $\mathrm{Tr}(\mathcal{W} V^T V) = 1$. Hence, we define the \emph{normalisation term}
\begin{equation}
\mathcal{N}\equiv \mathrm{Tr}(\mathcal{W} V^T V) = \sum_{j=1}^l w_j \abs{\boldsymbol{f}_j}^2
\label{normterm}
\end{equation} 
satisfying that $\mathcal{N} > 0$, with $\mathcal{N} = 1$ for normalised linear functions.

As for the second quantity, we can rewrite it as
\begin{align}
\mathrm{Tr}\left(\mathcal{W} V^T \mathcal{X}V\right) &= \mathrm{Tr}\left[\mathcal{W} V^T \left(\mathcal{I}-\mathbb{I}\right)V\right] = 
- \mathcal{N} + \sum_{i, j = 1}^l\sum_{k, m =1}^d \left(\mathcal{W}\right)_{ij} V_{kj} \mathcal{I}_{km} V_{mi}
\nonumber \\
&= - \mathcal{N} + \sum_{j = 1}^l w_j\sum_{k, m =1}^d V_{kj} 1_k 1_m V_{mj} = - \mathcal{N} + \sum_{j = 1}^l w_j \left(\sum_{k =1}^d V_{kj} 1_k \right)^2 
\nonumber \\
&= - \mathcal{N} + \sum_{j = 1}^l w_j \left(\boldsymbol{f}_j^\transpose\boldsymbol{1}\right)^2 = - \mathcal{N} + d \sum_{j = 1}^l w_j \abs{\boldsymbol{f}_j}^2 \mathrm{cos}^2\left(\varphi_{\boldsymbol{1},j}\right)
\nonumber \\
&= \sum_{j = 1}^l w_j \abs{\boldsymbol{f}_j}^2 \left[d\hspace{0.2em}\mathrm{cos}^2\left(\varphi_{\boldsymbol{1},j}\right) -1\right],
\label{geometrycalc}
\end{align}
where $\varphi_{\boldsymbol{1},j}$ is the angle between the vector associated with the $j$-th function and the direction defined by the vector of ones $\boldsymbol{1}$, and we have used the fact that $\abs{\boldsymbol{1}} = \sqrt{d}$.

Recalling that $|\mathrm{cos}\left(\varphi_{\boldsymbol{1},j}\right)| \leqslant 1$ and using equation (\ref{geometrycalc}), we see that $\mathrm{Tr}\left(\mathcal{W} V^T \mathcal{X}V\right)$ is bounded as
\begin{equation}
-\mathcal{N} \leqslant \mathrm{Tr}\left(\mathcal{W} V^T \mathcal{X}V\right) \leqslant \mathcal{N}(d-1),
\label{geometryinterval}
\end{equation}
and that the extremes are realised when either the functions are aligned with the direction of the vector of ones $\boldsymbol{1}$, or they lie in a subspace orthogonal to it and of dimension $(l-1)$. So, for sensor-symmetric networks with properties modelled by linear functions, there are two kinds of global properties that play a special role: the sum of all the natural parameters with equal weights, and any linear combination of them such that the sum of its coefficients vanishes. Any other set of global properties will produce some value for $\mathrm{Tr}\left(\mathcal{W} V^T \mathcal{X}V\right)$ lying within the interval in equation (\ref{geometryinterval}), and this will be given by the geometry of the transformation defined by $V \mathcal{W} V^T$. This motivates the introduction of the \emph{geometry parameter}
\begin{equation}
\mathcal{G} \equiv \frac{1}{\mathcal{N}}\mathrm{Tr}\left(\mathcal{W} V^T \mathcal{X}V\right) = \frac{1}{\mathcal{N}}\sum_{j = 1}^l w_j \abs{\boldsymbol{f}_j}^2 \left[d\hspace{0.2em}\mathrm{cos}^2\left(\varphi_{\boldsymbol{1},j}\right) -1\right],
\label{geometryparameter}
\end{equation}
which satisfies that $-1 \leqslant \mathcal{G} \leqslant (d-1)$. 

\begin{figure}[t]
\centering
\includegraphics[trim={0cm 0.1cm 1.2cm 0cm},clip,width=7.75cm]{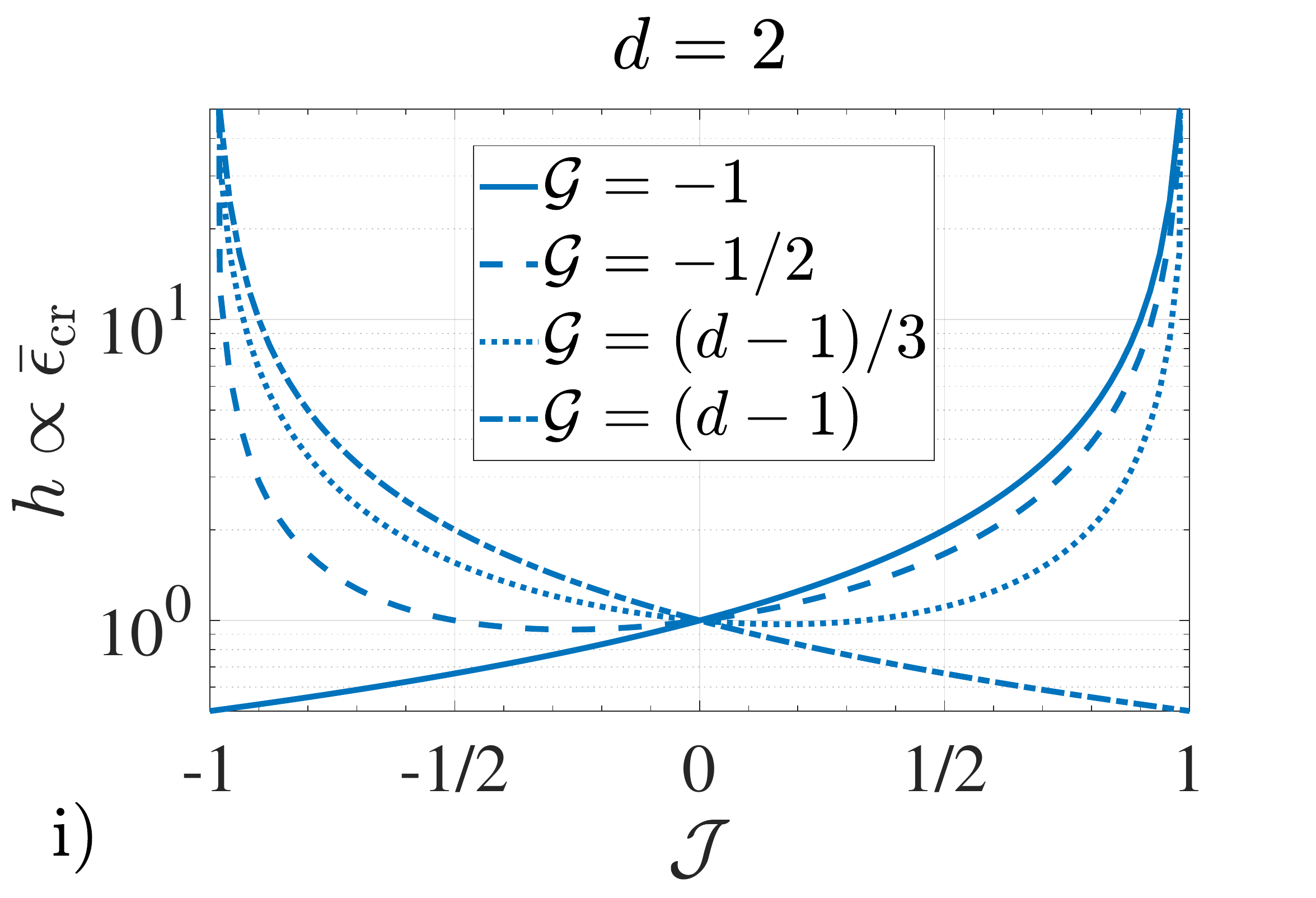}\includegraphics[trim={0cm 0.1cm 1.2cm 0cm},clip,width=7.75cm]{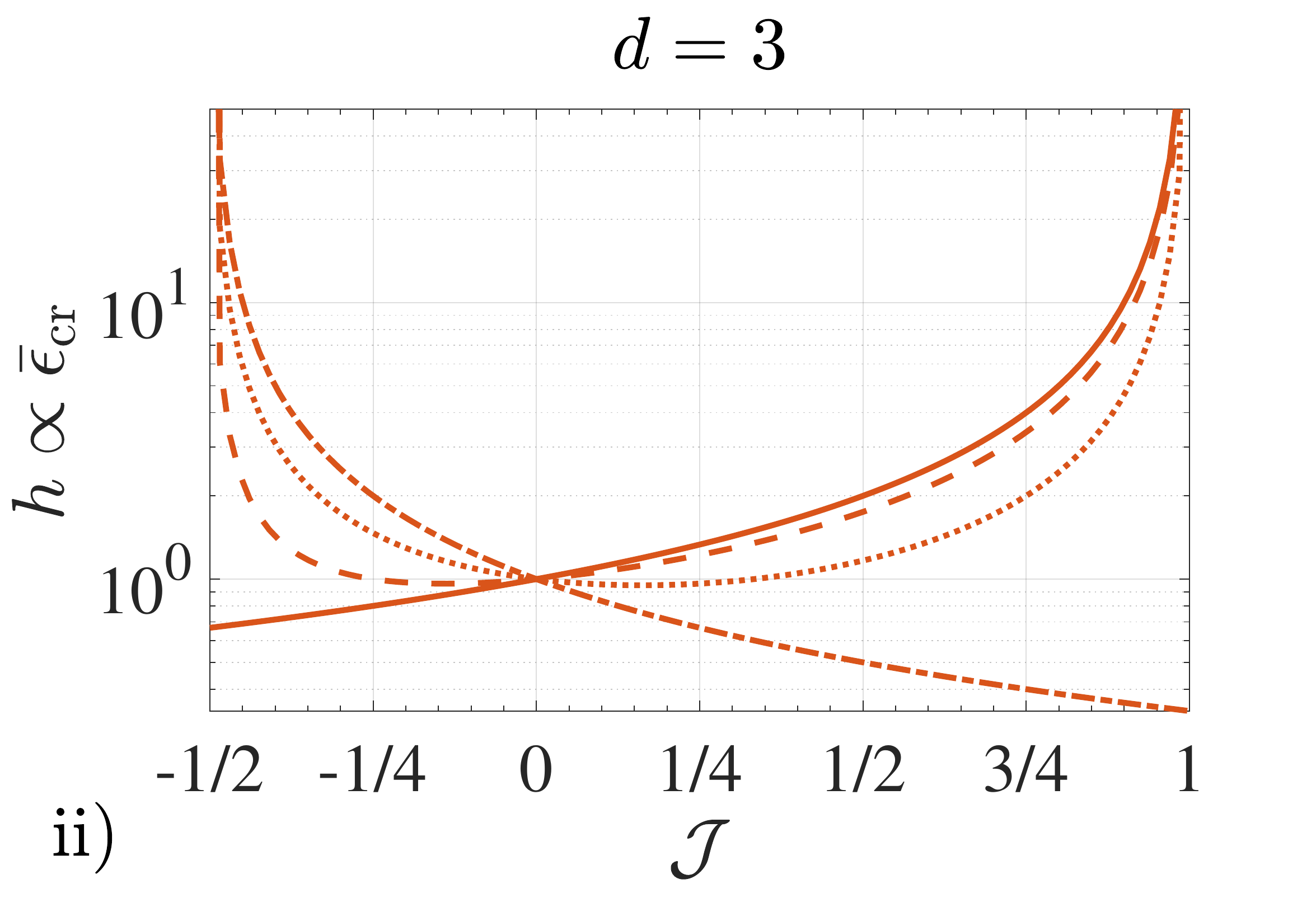}
\includegraphics[trim={0cm 0.1cm 1.2cm 0cm},clip,width=7.75cm]{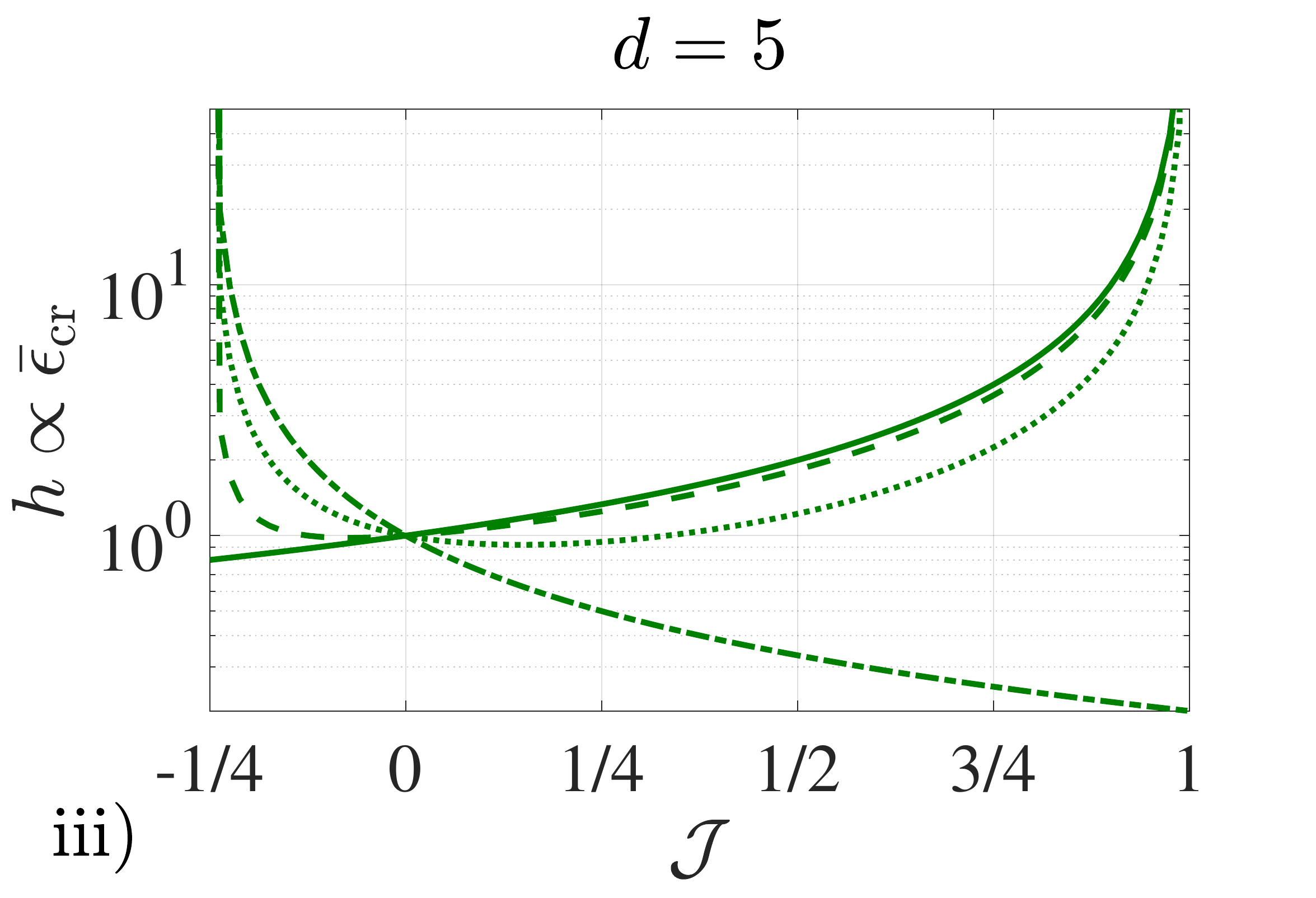}\includegraphics[trim={0cm 0.1cm 1.2cm 0cm},clip,width=7.75cm]{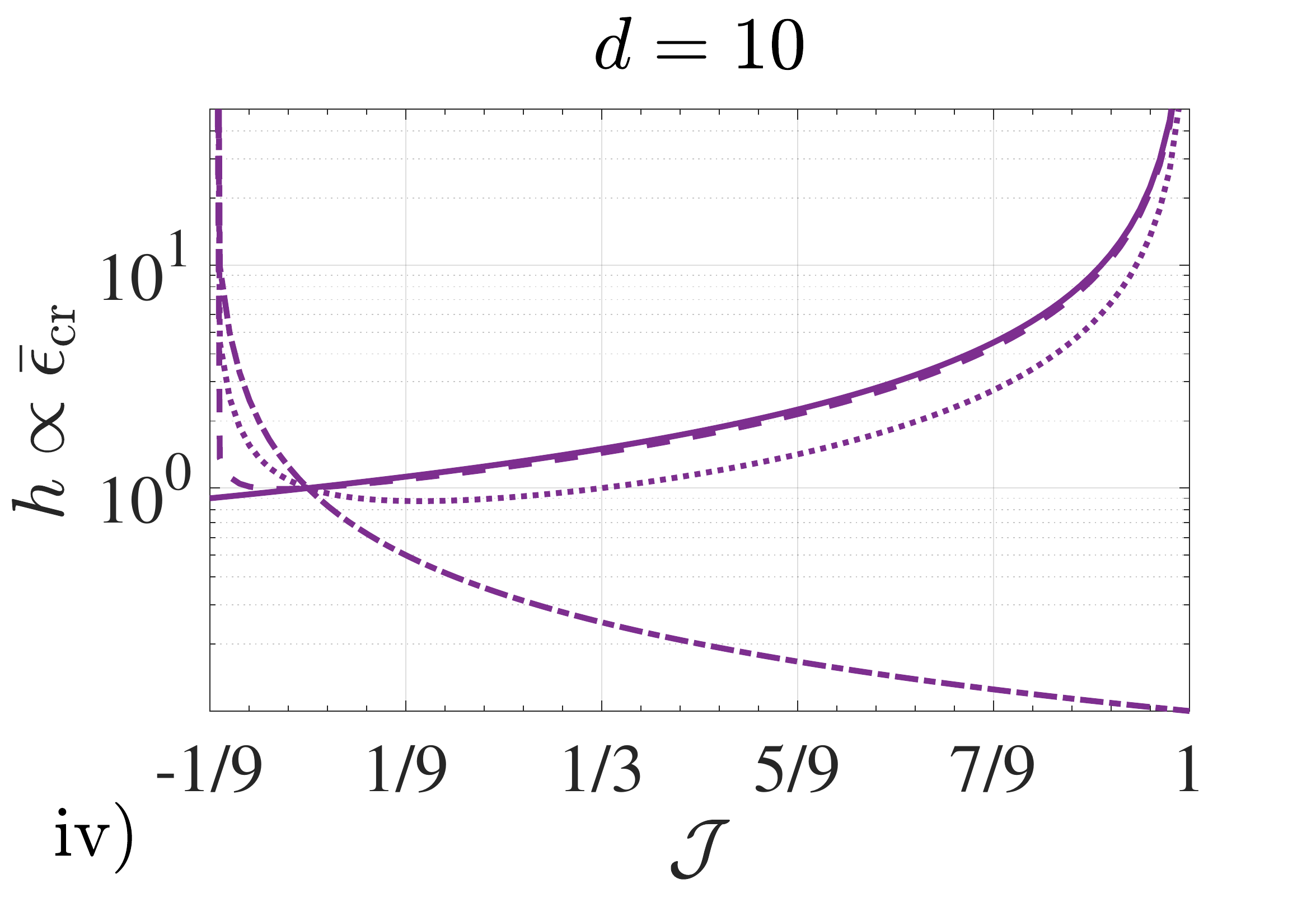}
	\caption{Representation of the interplay between the correlation strength $\mathcal{J}$ and the geometry parameter $\mathcal{G}$ in equation (\ref{geometrylinkfactor}) for a quantum sensing network with (i) $d=2$, (ii) $d=3$, (iii) $d=5$ and (iv) $d=10$ parameters. We observe that, given $\mathcal{G} \in (-1, (d-1))$, the minimum asymptotic uncertainty is achieved using a scheme with inter-sensor correlations of strength $\mathcal{J} \in (1/(1-d),1)$. The quantitative characterisation of these minima is provided in section \ref{subsec:intersensorasymp}.}
\label{geolinkplot}
\end{figure}

Inserting equations (\ref{normterm}) and (\ref{geometryparameter}) in equation (\ref{symmetricfunctionsfirst}), the asymptotic uncertainty finally becomes
\begin{equation}
\bar{\epsilon}_{\mathrm{cr}} = \frac{\mathcal{N}}{4\mu v}h\left(\mathcal{J}, \mathcal{G}, d\right),
\label{symmetricfunctionssecond}
\end{equation}
where
\begin{equation}
h\left(\mathcal{J}, \mathcal{G}, d\right) = \frac{\left[1 + (d-2 - \mathcal{G})\mathcal{J}\right]}{(1 -\mathcal{J})[1+(d-1)\mathcal{J}]}.
\label{geometrylinkfactor}
\end{equation}
Given a sensor-symmetric network with $d$ local properties, the factor $h\left(\mathcal{J}, \mathcal{G}, d\right)$ in equation (\ref{geometrylinkfactor}) codifies the interplay between the inter-sensor correlations of strength $\mathcal{J}$ and the geometry parameter $\mathcal{G}$ for any linear property, which may be local or global. A representation of this interplay can be found in figure \ref{geolinkplot}. The formulas in  equations (\ref{symmetricfunctionssecond}) and (\ref{geometrylinkfactor}) have been obtained without imposing further restrictions on the functions, and this implies that this formalism can be applied to any number of linear functions whose coefficients generate vectors that can form any angle and have any length. 

\subsection{The role of inter-sensor correlations I}
\label{subsec:intersensorasymp}

Let us exploit the previous result to address the problem of selecting a sensor-symmetric network state that is optimal to estimate a given set of linear functions. This amounts to finding the values for $v$ and $\mathcal{J}$ that are optimal for a given $\mathcal{G}$. One approach is to use the fact that, for qubits, $0\leqslant 4v\leqslant 1$, which allows us to lower bound equation (\ref{symmetricfunctionssecond}) as 
\begin{equation}
\bar{\epsilon}_{\mathrm{cr}} \geqslant \frac{\mathcal{N}}{\mu}h\left(\mathcal{J}, \mathcal{G}, d\right) \equiv \bar{\epsilon}_{\mathrm{f}}.
\end{equation}
We then search for the $\mathcal{J}$ that minimises this bound after having fixed $\mathcal{G}$, $d$ and $\mu$. In principle, there is no guarantee that the pairs of values $(4v = 1, \mathcal{J})$ generated by this method will correspond to any physical state, although the bounds on the asymptotic error constructed in this way would still be valid. Nevertheless, later we will study an example that realises a large portion of the pairs $(4v = 1, \mathcal{J})$ that we will predict.

\begin{figure}[t]
\centering
\includegraphics[trim={1cm 0.1cm 1.5cm 0.5cm},clip,width=14.75cm]{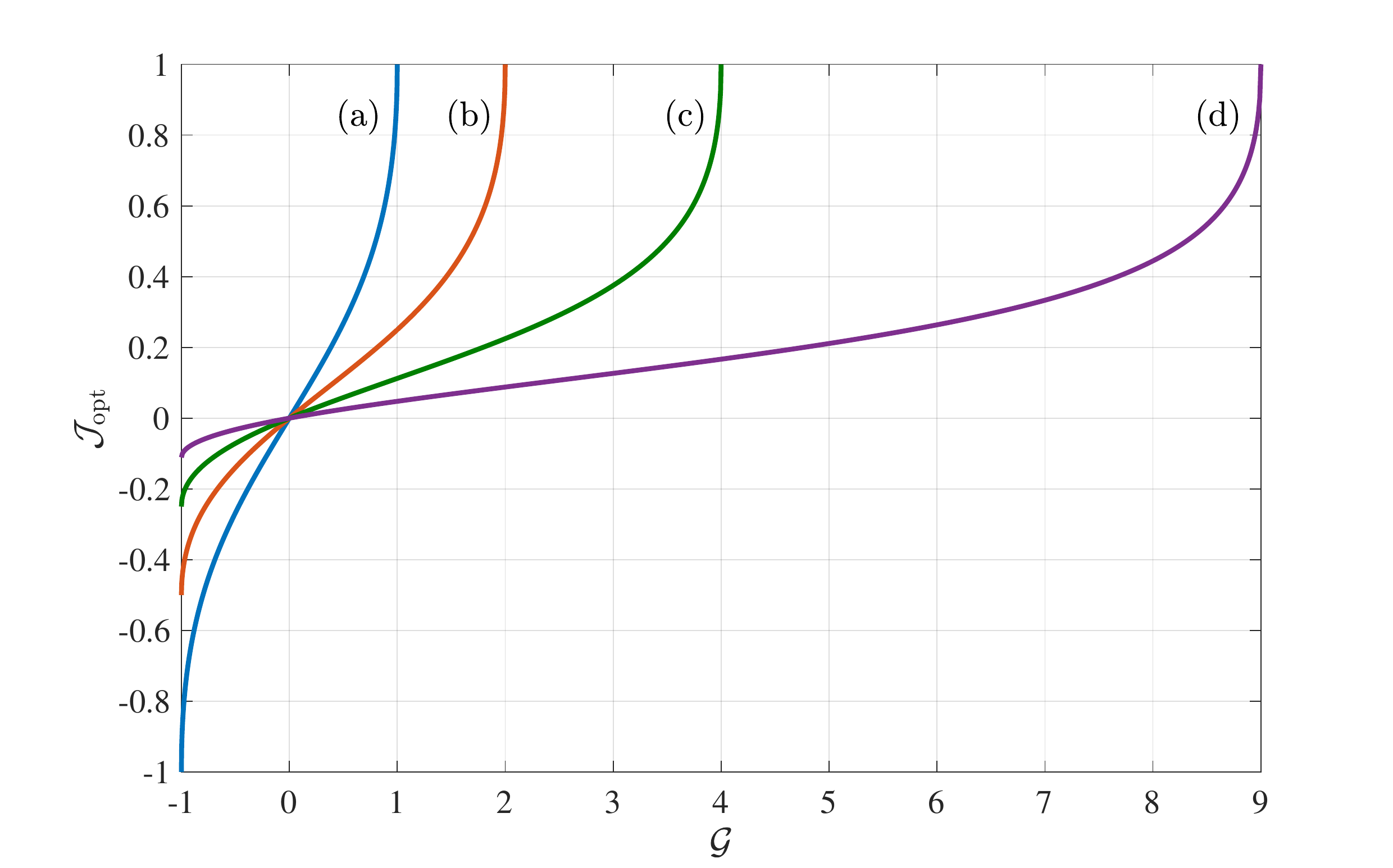}
	\caption{Optimal inter-sensor correlation strength $\mathcal{J}_{\mathrm{opt}}$ versus the geometry $\mathcal{G}$ of a set of arbitrary linear functions, for $d=2$, $3$, $5$ and $10$ parameters (lines (a - d) respectively). These monotonic curves provide a quantitative representation for the uncertainty minima identified in figure \ref{geolinkplot}, and the associated analytical formula is in equation (\ref{optgeolinkanalytical}). This result shows that, the more a collection of functions is clustered around the direction of $\boldsymbol{1}$, so that $\mathcal{G} = d - 1$, the larger the amount of correlations must be in order to perform the estimation optimally (provided that $4v=1$), while the opposite is true if the functions are instead clustered around the subspace orthogonal to $\boldsymbol{1}$, for which $\mathcal{G} = -1$. Remarkably, any amount of correlations is detrimental when $\mathcal{G} = 0$, even though a vanishing geometry parameter can \emph{also} be obtained for properties of the network that are global.}
\label{linkgeoent}
\end{figure}

By minimising $\bar{\epsilon}_{\mathrm{f}}$ (see \ref{asymlinearmin}) we find that, if $4v = 1$, and restricting our attention to the range $1/(1-d)<\mathcal{J} < 1$ where the information matrix is invertible, the optimal strength for the inter-sensor correlations of the network is
\begin{equation}
\mathcal{J}_{\mathrm{opt}} = \frac{1}{\mathcal{G}+2-d}\left[1- \sqrt{\frac{(\mathcal{G}+1)(d-1-\mathcal{G})}{d-1}}\right],
\label{optgeolinkanalytical}
\end{equation}
for $-1 < \mathcal{G} < d-1$, which is determined by the structure of the functions alone via $\mathcal{G}$ (once $d$ has been fixed). This provides a map between correlation strength and geometry with one-to-one correspondence (note that $\mathcal{J}_{\mathrm{opt}} \rightarrow (d-2)/[2(d-1)] $ when $\mathcal{G} \rightarrow d - 2$), as is illustrated in figure \ref{linkgeoent}, and this is the central result of our asymptotic analysis.

The expression in equation (\ref{optgeolinkanalytical}) reveals that, the more a collection of functions is clustered around the vector of ones $\boldsymbol{1}$, the larger the amount of positive correlations is required to be in order to perform the estimation optimally (provided that $4v=1$). Similarly, the amount of correlations with negative strength needs to be large if the functions are instead clustered around the subspace orthogonal to $\boldsymbol{1}$. The potential existence of this type of connection between geometry and quantum correlations was precisely one of the general open questions identifed in \cite{proctor2017networked}.

Furthermore, equation (\ref{optgeolinkanalytical}) (and figure \ref{linkgeoent}) shows that any non-zero pairwise correlation strength is detrimental whenever the geometry parameter vanishes. It is therefore interesting to investigate which kind of linear functions imply that $\mathcal{G} = 0$, as well as the form of the associated optimal strategy. To achieve this, let us recall the original definition for $\mathcal{G}$ in equation (\ref{geometryparameter}), that is, $\mathcal{G} = \mathrm{Tr}(\mathcal{W} V^\transpose \mathcal{X} V)/\mathcal{N}$. If we choose the uniform weighting matrix $\mathcal{W} = \mathbb{I}/l$, and if $V$ is an orthogonal transformation (i.e., $VV^\transpose = V^\transpose V =  \mathbb{I}$), then 
\begin{equation}
\mathcal{G} = \frac{1}{\mathcal{N} l }\mathrm{Tr}(V V^\transpose \mathcal{X}) = \frac{1}{\mathcal{N} l}\mathrm{Tr}(\mathcal{X})=\frac{1}{\mathcal{N} l}  \mathrm{Tr}(\mathcal{I} - \mathbb{I})= 0.
\end{equation}
Now we observe that $\mathcal{J} = 0$, which is the optimal choice for the previous scenario, is always achieved by a separable qubit state $\ket{\psi_0} = (\sqrt{a}\ket{0} + \sqrt{1 - a}\ket{1})^{\otimes d}$, and by selecting $a = 1/2$ we have that $4v = 1$. Thus we can say that the estimation of a set of $l = d$ linear functions that are equally relevant and orthogonal can be carried out optimally by preparing our scheme with separable states. Moreover, since the estimation of the parameters $\boldsymbol{\theta}$ is equivalent to choosing $V = \mathbb{I}$, our result implies that separable states are also optimal in that case. So, our present formalism is consistent with previous results \cite{proctor2017networked, proctor2017networkedshort, altenburg2018, kok2017}. 

The above conclusion is sufficient to affirm that while entangled pure states are generally useful for the optimal estimation of global properties, it is not true that we \emph{always} need entangled probes in such case. However, a transformation that is orthogonal preserves angles and lengths, and thus one may argue that, in a sense, the information encoded by a set of functions that gives rise to an orthogonal transformation is equivalent to the information content of the original parameters, provided that the weighting matrices are uniform. Hence, it is perhaps not surprising that a local estimation strategy is preferred here, since \cite{proctor2017networked, proctor2017networkedshort} had already shown that the estimation of local properties associated with commuting generators can be performed optimally with a local scheme. In view of this, it is important to establish whether there are other global properties with $\mathcal{G} = 0$ that instead select information that is not equivalent to estimating all the original parameters. First we observe that the eigendecomposition of $\mathcal{X}$, which is a symmetric matrix, is (see \ref{eigencalx})
\begin{equation}
\mathcal{X}_D = U_{\mathcal{X}}^\transpose \mathcal{X} U_{\mathcal{X}} = \mathrm{diag}\left[(d-1), -1, \dots, -1 \right], 
\end{equation} 
where the eigenvector for the first eigenvalue is $\boldsymbol{1}$ and those for the other eigenvalues belong to the orthogonal subspace. That implies that if we choose a single linear function as $V = \boldsymbol{f} = U_{\mathcal{X}}\boldsymbol{1}$, then we will have that $\mathcal{G} = \boldsymbol{1}^\transpose U_{\mathcal{X}}^\transpose \mathcal{X} U_{\mathcal{X}} \boldsymbol{1} /d= \boldsymbol{1}^\transpose \mathcal{X}_D \boldsymbol{1} /d =0$. Now consider a three-parameter network, so that 
\begin{align}
\boldsymbol{f} = U_{\mathcal{X}}\boldsymbol{1} = \frac{1}{\sqrt{6}}
\begin{pmatrix}
\sqrt{2} & \sqrt{3} & 1\\
\sqrt{2} & -\sqrt{3} & 1\\
\sqrt{2} & 0 & -2\\
\end{pmatrix} 
\begin{pmatrix}
1 \\
1 \\
1 \\
\end{pmatrix} 
= \frac{1}{\sqrt{6}}
\begin{pmatrix}
\sqrt{2} + \sqrt{3} + 1 \\
\sqrt{2} - \sqrt{3} + 1 \\
\sqrt{2} -2 \\
\end{pmatrix}.
\end{align}
Clearly,  this gives rise to a global property, as these are the coefficients of a non-trivial function of three local parameters. Yet, $\mathcal{G} = 0$, and so, according to equation (\ref{optgeolinkanalytical}), pairwise correlations are detrimental. Therefore, entanglement is sometimes not needed in scenarios where we are estimating non-trivial global properties. Interestingly, the same argument fails for $d=2$, since in that case
\begin{equation}
\boldsymbol{f} = U_{\mathcal{X}}\boldsymbol{1} = \frac{1}{\sqrt{2}}
\begin{pmatrix}
1 & 1 \\
1 & -1
\end{pmatrix}
\begin{pmatrix}
1 \\
1
\end{pmatrix} =
\begin{pmatrix}
\sqrt{2} \\
0
\end{pmatrix},
\end{equation}
and this is associated with a local property because it simply rescales the first parameter. Nonetheless, our conclusion above is still valid in general.

For the link between geometry and correlations in equation (\ref{optgeolinkanalytical}) to be truly relevant, it is necessary that there are physical states with the properties that such a link predicts as optimal. In \cite{proctor2017networked} we studied the estimation of $1 \leqslant l \leqslant d = 2$ linear and normalised but otherwise arbitrary functions using the sensor-symmetric state 
\begin{equation}
\ket{\psi_0} = \frac{1}{\sqrt{2\left(1+\gamma^2\right)}}\left[\ket{0 0} + \gamma\left(\ket{01} + \ket{10}  \right) + \ket{1 1}\right],
\label{gammastatetwo}
\end{equation}
with $-\infty < \gamma < \infty$, and we provided a complete solution to this two-parameter estimation problem. The fact that this is a particular case of the more general formalism that we develop in this work suggests that, for the $d=2$ case, it may be possible to use the state in  equation (\ref{gammastatetwo}) to realise all the pairs $(4v = 1, \mathcal{J})$ that are optimal according to our results. We will now show that this is the case. 

Recalling that $\sigma_z \ket{i} = (-1)^{i}\ket{i}$, we see that, for the state in equation (\ref{gammastatetwo}), $\langle \sigma_{z,1} \rangle = \langle \sigma_{z,2} \rangle = 0$ and $\langle \sigma_{z,1} \sigma_{z,2} \rangle = \langle \sigma_{z,1} \sigma_{z,2} \rangle = (1-\gamma^2)/(1+\gamma^2)$, so that the variance is $4v = 4v_1 = 4v_2 = 1$ and the quantifier for the inter-sensor correlations can be written as a function of $\gamma$ as $\mathcal{J}=(1-\gamma^2)/(1+\gamma^2)$. This function reaches the maximum $\mathcal{J}=1$ at $\gamma = 0$, while it tends monotonically from such point to $\mathcal{J} = - 1$ when $\gamma \rightarrow \pm \infty$. In other words, for $d=2$ there is always a physical state that satisfies the condition imposed in equation (\ref{optgeolinkanalytical}) when $4v = 1$.

It is interesting to observe that $\gamma$ splits the state into a part where the sum of the parameters is encoded and a part that encodes the difference. More concretely,
\begin{align}
\mathrm{e}^{-\frac{i}{2}(\sigma_{z,1}\theta_1+\sigma_{z,2}\theta_2)}\ket{\psi_0} = &\frac{1}{\sqrt{2\left(1+\gamma^2\right)}}\left[\mathrm{e}^{-\frac{i}{2}(\theta_1 + \theta_2)}\ket{0 0} + \mathrm{e}^{\frac{i}{2}(\theta_1 + \theta_2)}\ket{1 1} 
\right] 
\nonumber \\
&+ \frac{\gamma}{\sqrt{2\left(1+\gamma^2\right)}}\left[\mathrm{e}^{-\frac{i}{2}(\theta_1 - \theta_2)}\ket{01} + \mathrm{e}^{\frac{i}{2}(\theta_1 - \theta_2)}\ket{10} \right].
\label{twonetworktransformed}
\end{align}
A partial extension of this idea to the $d$-parameter case can be achieved by constructing a state where the part that encodes functions aligned with the direction of $\boldsymbol{1}$ is isolated in an analogous fashion, i.e., 
\begin{align}
\ket{\psi_0} &= \frac{1}{\sqrt{2\left[1 + \left( 2^{d-1}-1 \right)\gamma^2 \right]}} \left[ \ket{0 0 \dots 0} + \ket{1 1 \dots 1} + \gamma \left(\text{all other terms}\right) \right]
\nonumber \\
&= \frac{1}{\sqrt{2\left[1 + \left( 2^{d-1}-1 \right)\gamma^2 \right]}} \left[\left(1-\gamma\right)\left(\ket{0}^{\otimes d} + \ket{1}^{\otimes d}\right) + \gamma \left(\ket{0} + \ket{1} \right)^{\otimes d} \right].
\label{gammastategen}
\end{align}
For this probe, $4v_i = 1 - \langle \sigma_{z,i} \rangle^2 = 1  = 4v$ for all $i$, and $4c_{ij} = \langle \sigma_{z,i} \sigma_{z,j} \rangle - \langle \sigma_{z,i} \rangle\langle \sigma_{z,j} \rangle = \langle \sigma_{z,i} \sigma_{z,j} \rangle = (1-\gamma^2)/[1 + (2^{d-1}-1)\gamma^2] = 4c$ for all $i\neq j$, which verifies that the state in equation (\ref{gammastategen}) is also sensor symmetric. As a result, we can see that its inter-sensor correlations are given by
\begin{equation}
\mathcal{J} = \frac{1-\gamma^2}{1 + \left(2^{d-1}-1\right)\gamma^2}.
\label{gengammacorrelations}
\end{equation} 
If $0 \leqslant \abs{\gamma} \leqslant 1$, then we have that $1 \geqslant \mathcal{J} \geqslant 0$. This implies that there always exists a physical state associated with all the results in this section that require either positive inter-sensor correlations, or the absence of them. On the other hand, the amount of negative correlations that this state can cover lies in $ 0 > \mathcal{J} > - 1/(2^{d-1}-1)$, which corresponds to $1 < \abs{\gamma} < \infty$. Unfortunately, the amount of negative correlations that equation (\ref{optgeolinkanalytical}) might predict can lie in $ 0 > \mathcal{J} > 1/(1-d)$, where $1/(1-d) \leqslant - 1/(2^{d-1}-1)$ for $d\geqslant 2$ and the inequality is only saturated when $d=2$. Thus there is a subinterval not covered by equation (\ref{gammastategen}). Whether there are other physical states that may realise the missing values is an open question. 

Finally, we note that the only entangled pure probes that may be asymptotically relevant for sensor-symmetric networks are those that give rise to inter-sensor correlations, while any other form of entanglement will be irrelevant in this type of scenario. To illustrate this idea, let us consider the state in equation (\ref{gammastategen}) for $d = 3$, and suppose that the functions to be estimated give rise to the geometry parameter $\mathcal{G} = 0$. We have seen that, in that case, no inter-sensor correlations are needed to perform the estimation optimally, which implies that, according to equation (\ref{gengammacorrelations}), $\gamma = \pm 1$ . By inserting these parameters in equation (\ref{gammastategen}) we find that the optimal states are
\begin{equation}
\ket{\psi_{+}} = \frac{1}{2\sqrt{2}}\left(\ket{0}+\ket{1}\right)^{\otimes 3}
\end{equation}
and
\begin{equation}
\ket{\psi_{-}} = \frac{1}{2\sqrt{2}} \left[2\left(\ket{0}^{\otimes 3} + \ket{1}^{\otimes 3}\right) - \left(\ket{0} + \ket{1} \right)^{\otimes 3} \right].
\end{equation}
The first state is separable, but $\ket{\psi_{-}}$ is not. More concretely, if we tried to write the latter as
$\ket{\psi_{-}} = (x_0\ket{0}+x_1\ket{1})(y_0\ket{0}+y_1\ket{1})(z_0\ket{0}+z_1\ket{1})$, with $|x_0|^2+|x_1|^2 = |y_0|^2+|y_1|^2  = |z_0|^2+|z_1|^2 = 1$, we would find contradictions such as 
\begin{equation}
\left[(x_0 = x_1) \land (x_0 = - x_1)\right]\land \left(|x_0|^2+|x_1|^2 = 1\right),
\end{equation}
which by \emph{reductio ad absurdum} allows us to conclude that the state with $\gamma = -1$ and $d = 3$ is entangled. Hence, while here entanglement is not required to reach the asymptotic optimum, neither is it necessarily detrimental. The only requirement imposed by our formalism is the absence of pairwise correlations, and the presence or absence of any other kind of correlation does not affect the asymptotic uncertainty. 

\subsection{Optimal POVM in the asymptotic regime}\label{asympovm}

The final step of the asymptotic analysis is to find some POVM that is optimal in the large-$\mu$ regime, in the sense that it saturates the quantum Cram\'{e}r-Rao bound as $\epsilon_{\mathrm{asym}}^c = \bar{\epsilon}_{\mathrm{cr}}$, and we can achieve this by requiring that $F(\boldsymbol{\theta})=F_q$ \cite{sammy2016compatibility, pezze2017simultaneous}. That the latter condition refers to the parameters but not to the functions, together with the fact that the former can be estimated optimally using a local strategy \cite{proctor2017networked, proctor2017networkedshort} (see also section \ref{subsec:intersensorasymp}), suggests that a local POVM might be sufficient to make the classical and quantum information matrices equal. In fact, this would be very useful, since then we could associate any enhancement derived from the presence of correlations with the initial state alone. In the following we demonstrate this for a network with $d=2$ parameters.

Consider a local POVM with elements 
\begin{equation}
\ket{n, k} = \left[\ket{0} + (-1)^n \ket{1}\right]\otimes[\ket{0} + (-1)^k \ket{1}]/2,
\label{localpovm}
\end{equation}
where $n, k = 0, 1$. Furthermore, we have seen that, if $d=2$, then the state in equation (\ref{gammastatetwo}) is general enough to realise all the asymptotic results predicted by our theory. As such, this is the probe that we will use in this calculation. Combining this POVM with the transformed state $\ket{\psi(\theta_1, \theta_1)} = \mathrm{e}^{-\frac{i}{2}(\sigma_{z,1}\theta_1+\sigma_{z,2}\theta_2)}\ket{\psi_0}$ in equation (\ref{twonetworktransformed}), we find the amplitude
\begin{align}
\braket{n,k}{\psi(\theta_1, \theta_2)} &\propto \mathrm{e}^{-\frac{i}{2}(\theta_1+\theta_2)} + (-1)^{n+k}\mathrm{e}^{\frac{i}{2}(\theta_1+\theta_2)}
+\gamma\left[(-1)^k \mathrm{e}^{-\frac{i}{2}(\theta_1-\theta_2)} + (-1)^n \mathrm{e}^{\frac{i}{2}(\theta_1-\theta_2)}\right]
\nonumber \\
& \propto \mathrm{cos}\left\lbrace\left[\theta_1 + \theta_2 + \pi(k+n)\right]/2\right\rbrace
+ \gamma\hspace{0.15em}\mathrm{cos}\left\lbrace\left[\theta_1 - \theta_2 - \pi(k-n)\right]/2\right\rbrace,
\end{align}
the modulus of the proportionality factor being $1/\sqrt{2(1+\gamma^2)}$. This allows us to arrive at the likelihood function
\begin{equation}
p(n, k | \theta_1, \theta_2) = ||\braket{n,k}{\psi(\theta_1, \theta_2)}||^2 = \left[\mathrm{cos}(x_{+}) + \gamma \mathrm{cos}(x_{-})\right]^2/[2(1+\gamma^2)],
\label{multilikelihood}
\end{equation}
where we have introduced the notation $x_{\pm} \equiv \left[\theta_1 \pm \theta_2 \pm \pi(k\pm n)\right]/2$.

The elements of the classical Fisher information matrix in equation (\ref{fim}) for the quantum probability in equation (\ref{multilikelihood}) are
\begin{align}
[F(\boldsymbol{\theta})]_{11} &= \sum_{n,k = 0}^1 \frac{1}{p(n, k | \theta_1, \theta_2)}\left[\frac{\partial p(n, k | \theta_1, \theta_2)}{\partial \theta_1} \right]^2 
\nonumber \\
&= \frac{1}{2\left(1+\gamma^2 \right)} \sum_{n,k = 0}^1  \left[ \mathrm{sin}(x_{+}) + \gamma \mathrm{sin}(x_{-}) \right]^2 = 1,
\label{classfimqubit1}
\end{align}
\begin{align}
[F(\boldsymbol{\theta})]_{22} &= \sum_{n,k = 0}^1 \frac{1}{p(n, k | \theta_1, \theta_2)}\left[\frac{\partial p(n, k | \theta_1, \theta_2)}{\partial \theta_2} \right]^2 
\nonumber \\
&= \frac{1}{2\left(1+\gamma^2 \right)} \sum_{n,k = 0}^1 \left[ \mathrm{sin}(x_{+}) - \gamma \mathrm{sin}(x_{-}) \right]^2 = 1,
\end{align}
and
\begin{align}
[F(\boldsymbol{\theta})]_{12} &= \sum_{n,k = 0}^1 \frac{1}{p(n, k | \theta_1, \theta_2)}\frac{\partial p(n, k | \theta_1, \theta_2)}{\partial \theta_1}\frac{\partial p(n, k | \theta_1, \theta_2)}{\partial \theta_2} 
\nonumber \\
&= \frac{1}{2\left(1+\gamma^2 \right)} \sum_{n,k = 0}^1 \left[ \mathrm{sin}^2(x_{+}) - \gamma^2 \mathrm{sin}^2(x_{-})  \right] = \frac{1-\gamma^2}{1+\gamma^2},
\label{classfimqubit3}
\end{align}
with $[F(\boldsymbol{\theta})]_{21}= [F(\boldsymbol{\theta})]_{12}$. Additionally, in sections \ref{sec:networksasym} and \ref{subsec:intersensorasymp} we have seen that, for this configuration,
\begin{equation}
F_q = 
\begin{pmatrix}
1 & \mathcal{J} \\
\mathcal{J} & 1
\end{pmatrix} =
\begin{pmatrix}
1 & (1-\gamma^2)/(1+\gamma^2) \\
(1-\gamma^2)/(1+\gamma^2) & 1
\end{pmatrix},
\end{equation}
which is identical to the classical Fisher information matrix in equations (\ref{classfimqubit1} - \ref{classfimqubit3}). We thus conclude that the quantum strategy formed by the local POVM in equation (\ref{localpovm}) and the state in equation (\ref{gammastatetwo}) is asymptotically optimal. This completes our solution for the asymptotic estimation of linear functions in a two-parameter network, and will be our starting point to perform a Bayesian analysis.

\section{Bayesian analysis of non-asymptotic quantum sensing networks}
\label{bayesianresults}

Now we turn to the more general problem of estimating linear functions when different amounts of data are available, which may include cases with a low number of trials. Thanks to the simplicity of the asymptotic approach, in section \ref{asymptoticresults} we were able to discuss examples where $d = 2, 3, 5$ and $10$, and many of the results there were valid for any $d$. However, due to the more challenging nature of the numerical calculations associated with Bayesian estimation, in the remainder of this work we will focus on \emph{two-parameter} sensor-symmetric qubit networks.

\subsection{Regions of unambiguous information}
\label{subsec:multiprioranalysis}

Our aim is to use the asymptotically optimal strategy in equations (\ref{twonetworktransformed}), (\ref{localpovm}) and (\ref{multilikelihood}) as a guide to perform a non-asymptotic analysis. Following our discussion in section \ref{hybrid}, this approach is best justified when, as $\mu$ grows, the likelihood function 
\begin{equation}
p(\boldsymbol{n}, \boldsymbol{k} |\theta_1, \theta_2) = \prod_{i=1}^\mu p(n_i, k_i|\theta_1, \theta_2),
\end{equation}
with each $p(n_i, k_i|\theta_1, \theta_2)$ given by equation (\ref{multilikelihood}), becomes concentrated around a \emph{unique} absolute maximum within the prior area $\Delta_0$. Indeed, this condition helps to prevent the estimation process from giving ambiguous answers \cite{jaynes2003}. Hence, before we proceed we need to find how large $\Delta_0$ can be such that the above requirement is satisfied.

One way of estimating this size is to first represent the posterior probability $p(\theta_1, \theta_2|\boldsymbol{n}, \boldsymbol{k}) \propto p(\boldsymbol{n}, \boldsymbol{k} |\theta_1, \theta_2)$ as a function of $(\theta_1, \theta_2)$, where the outcomes $(\boldsymbol{n}, \boldsymbol{k})$ come from a simulation with true values $(\theta_1', \theta_2')$, and then visualise the regions with an asymptotically unique absolute maximum in a direct fashion (see \cite{jesus2017}). 

\begin{figure}[t]
\centering
\includegraphics[trim={0.2cm 0cm 0.5cm 0cm},clip,width=5.1cm]{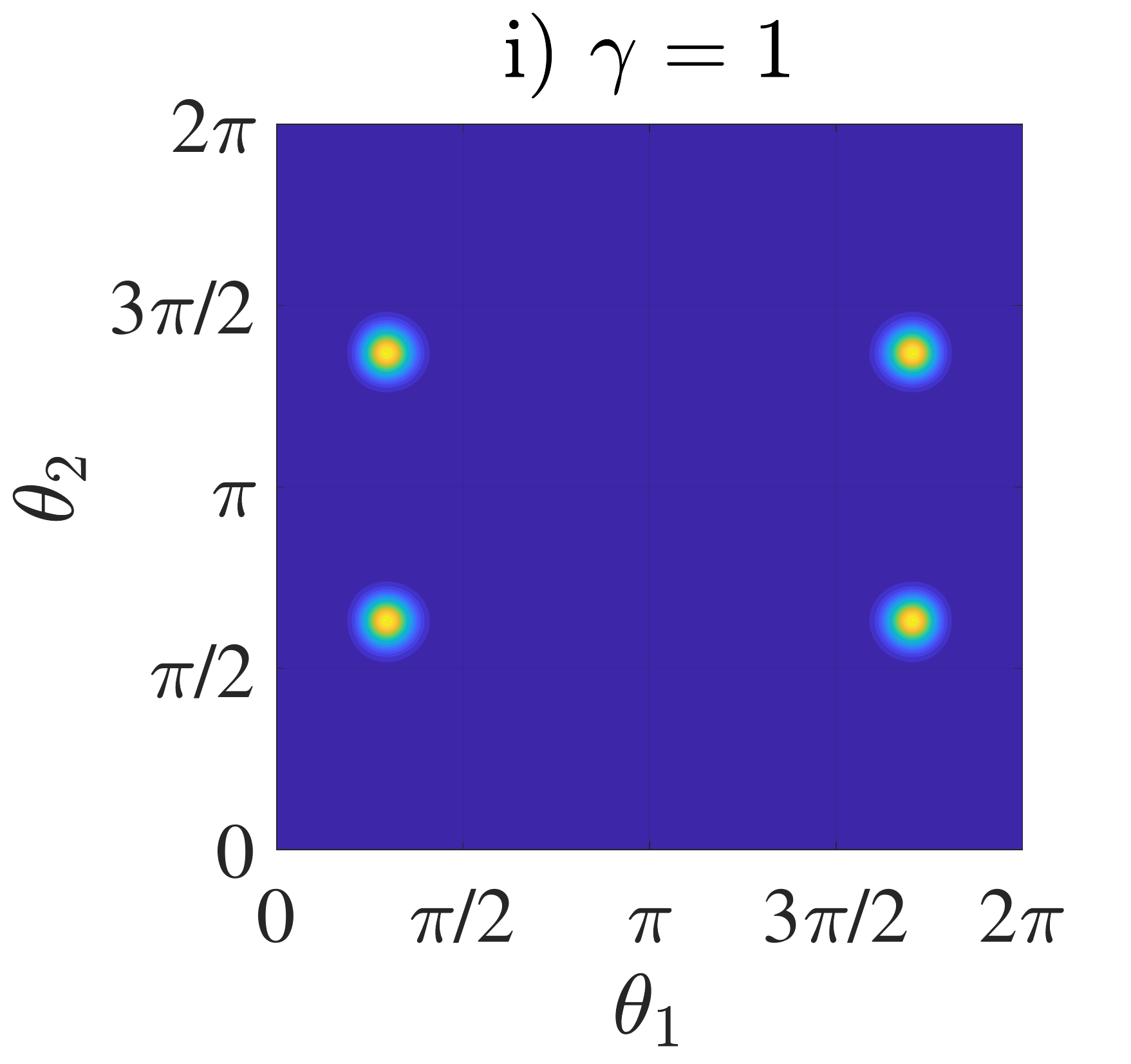}\includegraphics[trim={0.2cm 0cm 0.5cm 0cm},clip,width=4.3cm]{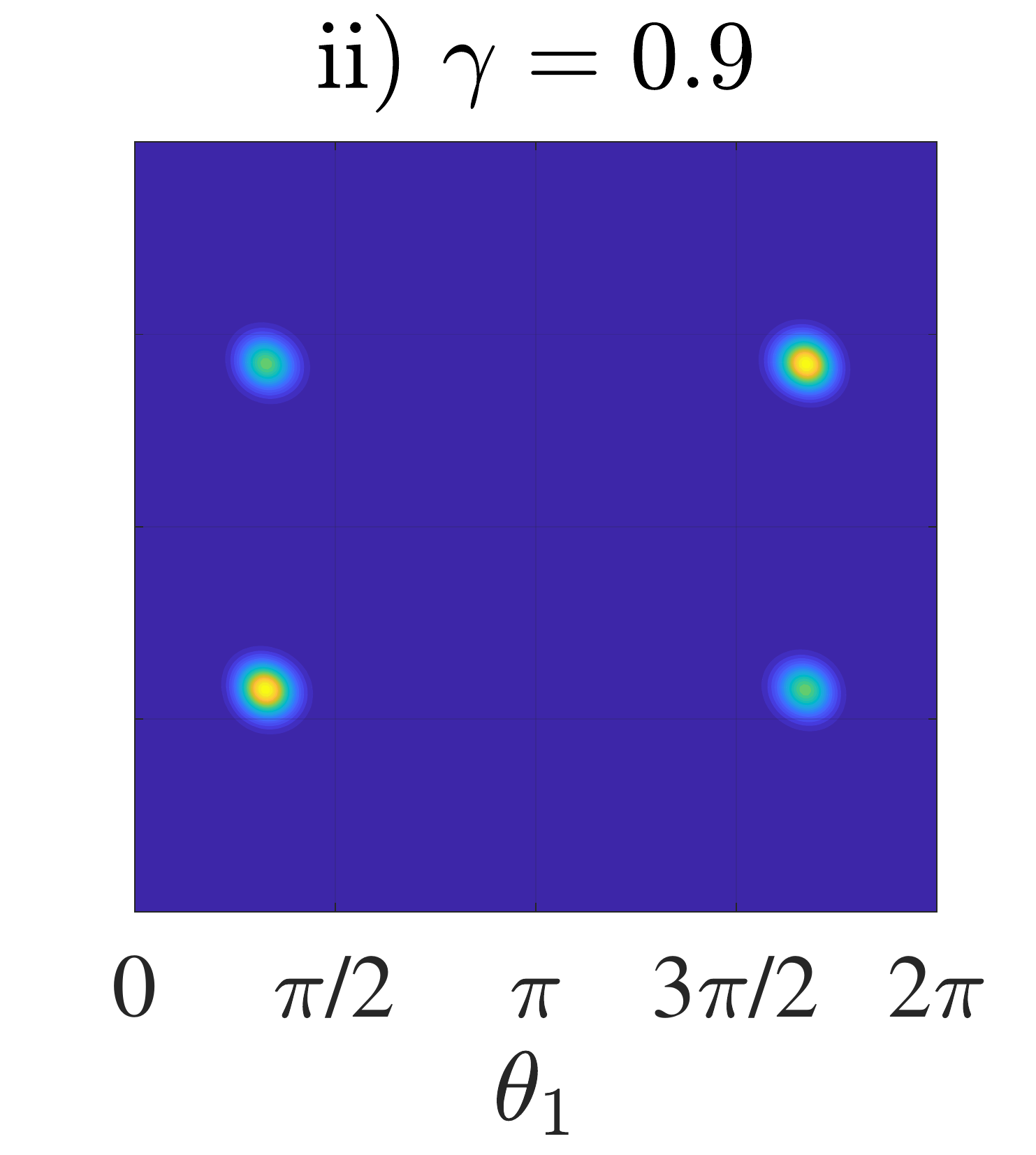}\includegraphics[trim={0.2cm 0cm 0.5cm 0cm},clip,width=4.3cm]{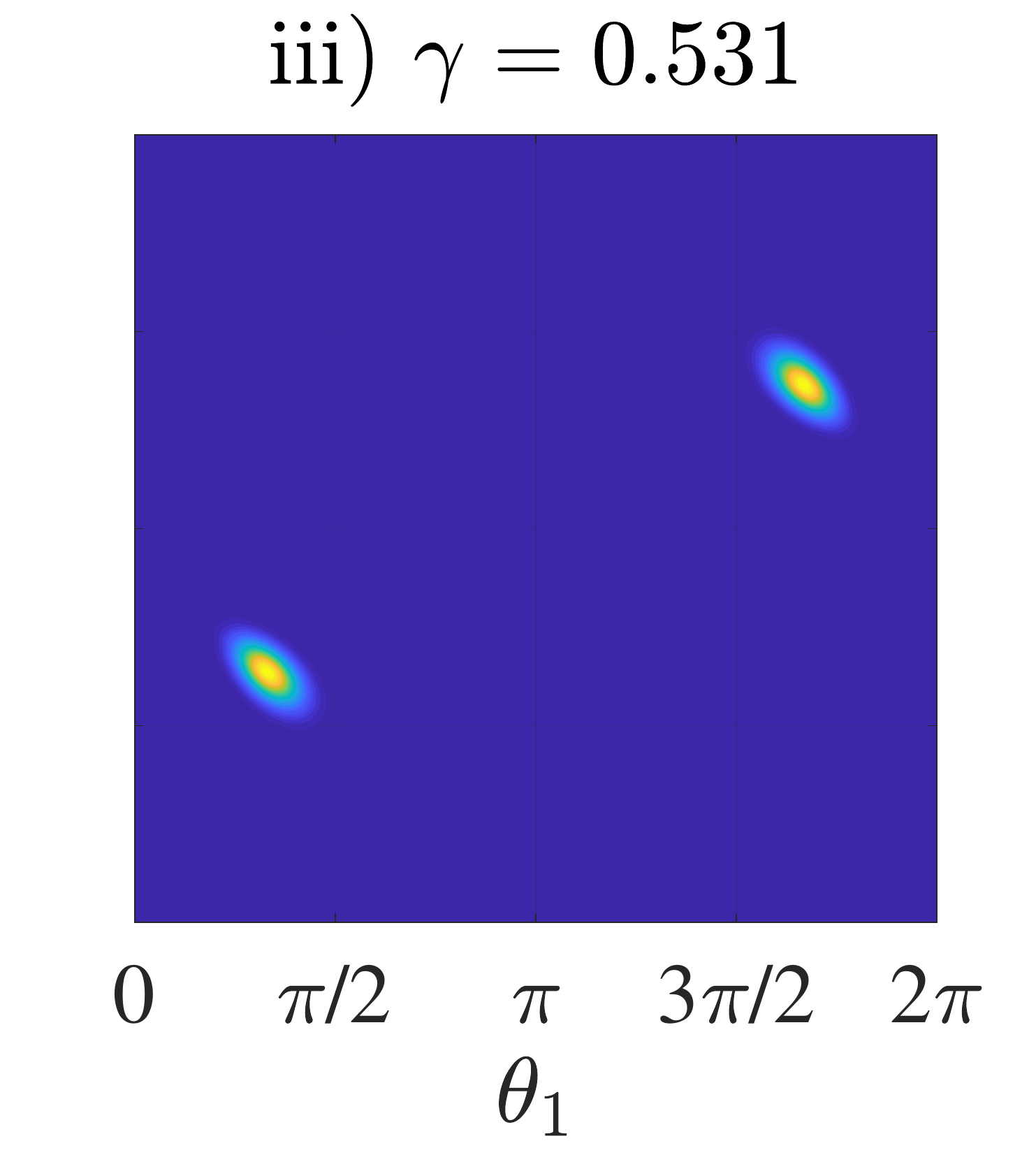}
\includegraphics[trim={0.2cm 0cm 0.5cm 0cm},clip,width=5.1cm]{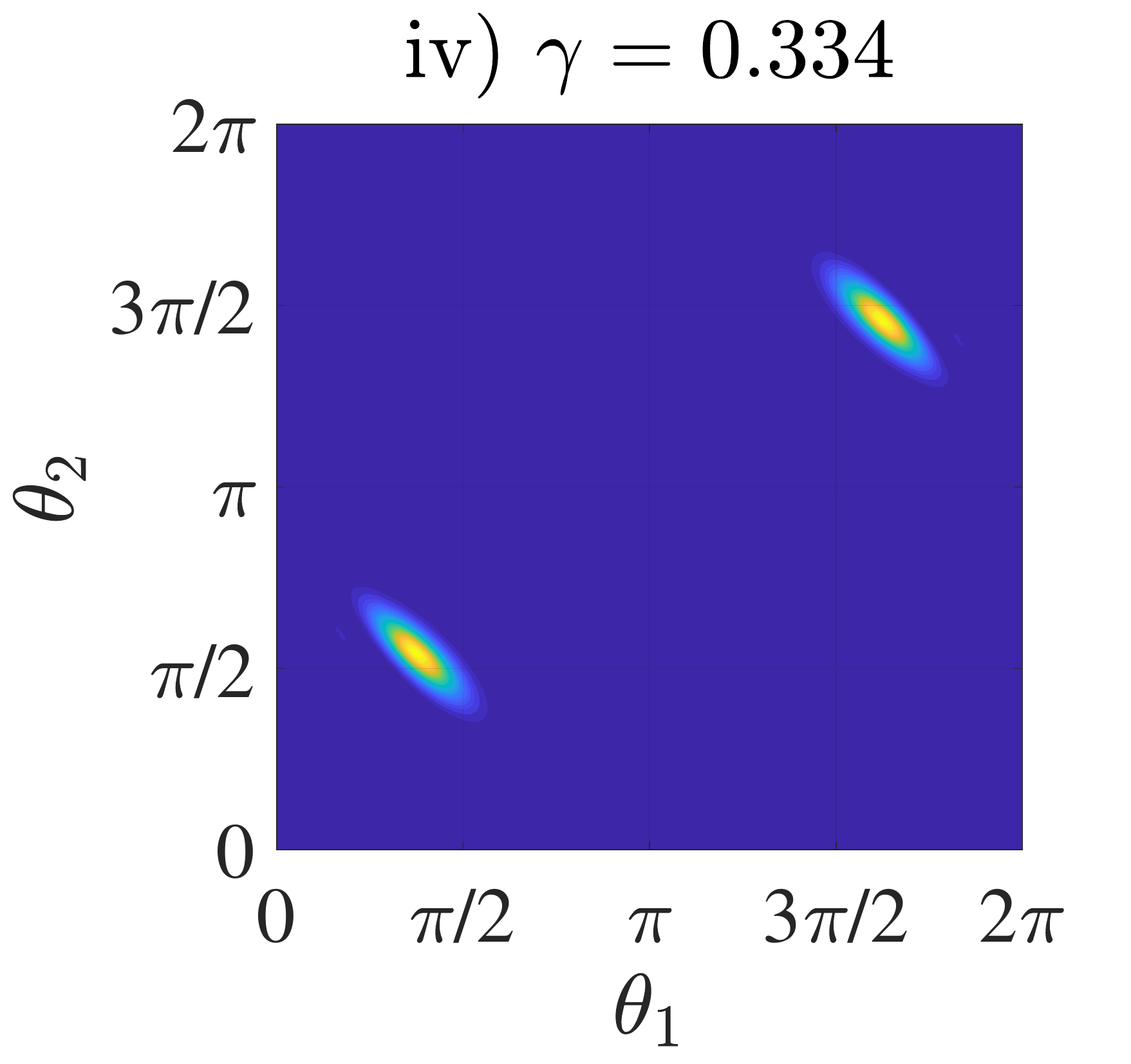}\includegraphics[trim={0.1cm 0cm 0.5cm 0cm},clip,width=4.35cm]{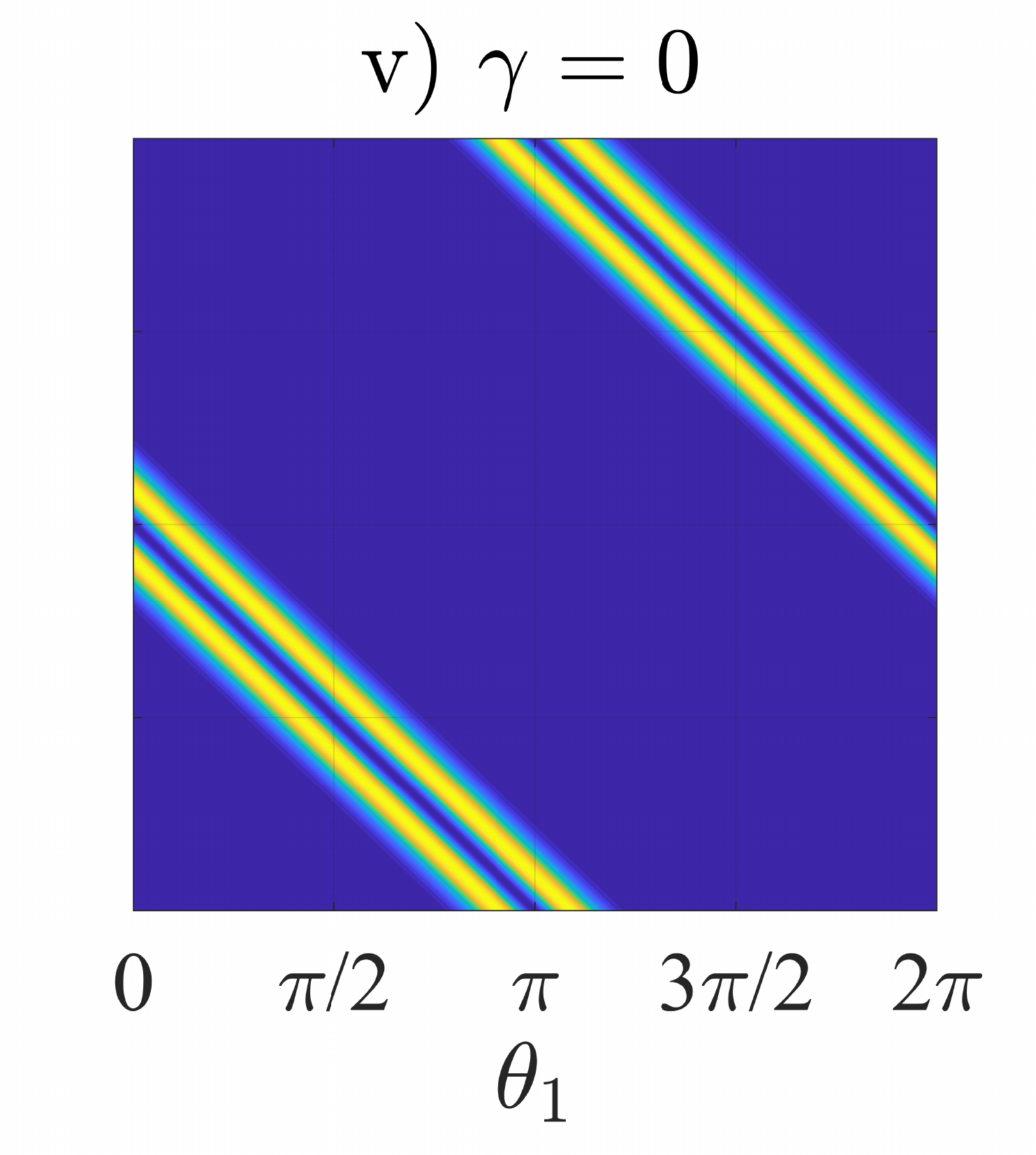}
\caption{Posterior density functions for random simulations of $\mu = 100$ trials, a flat prior and the quantum strategy represented by the likelihood in equation (\ref{multilikelihood}), with (i) $\gamma = 1$, (ii) $\gamma = 0.9$, (iii) $\gamma = 0.531$, (iv) $\gamma = 0.334$ and (v) $\gamma = 0$. The simulated true values of the parameters are $\theta'_1=1$ and $\theta'_2=2$. This figure shows that the potential ambiguities in the estimation associated with scenarios (i - iv) can be generally avoided if the prior area satisfies $\Delta_0 \leqslant \pi^2$. On the contrary, while the scheme (v) can be exploited to estimate the sum of the parameters, in general it cannot provide good estimates for other linear functions, independently of the value for $\Delta_0$. We draw attention to the fact that a similar pattern emerges as  $\gamma \rightarrow \infty$, but with the posterior peaks tending to the direction orthogonal to that in (v).}
\label{priornetwork}
\end{figure}

The previous method generates the results shown in figure \ref{priornetwork} for several values of $\gamma$. First we note that the simulations in figure \ref{priornetwork}  have been restricted to the area $(\theta_1, \theta_2) \in [0, 2\pi]\times [0, 2\pi]$ because the single-shot likelihood in equation (\ref{multilikelihood}) is invariant under $\theta_i \rightarrow \theta_i + 2\pi m$, with $m = 0, \pm 1, \pm 2, \dots$ and $i=1,2$, and thus it suffices to examine its symmetries within one period.  Depending on the value for $\gamma$, we see that the posterior probability in figures \ref{priornetwork}.i - \ref{priornetwork}.iv develops either two or four identical absolute maxima as $\mu$ grows, and that each of these peaks is located within an extension of area $\pi^2$. Therefore, in the presence of complete ignorance, i.e., $\Delta_0 = 4 \pi^2$, the quantum strategy under analysis cannot select a unique answer, a phenomenon already encountered in single-parameter metrology \cite{kolodynski2014, jesus2017, jesus2018, jesus2019thesis}. In view of this, to avoid the ambiguities in figures \ref{priornetwork}.i - \ref{priornetwork}.iv we shall impose that the prior area satisfies the condition $\Delta_0 \leqslant \pi^2$.

The situation for $\gamma = 0$ in figure \ref{priornetwork}.v is, however, different. In that case, no single peak can be selected even if $\mu \gg 1 $, which implies that such scheme does not have an asymptotic approximation in the sense of section \ref{hybrid}. This is consistent with the fact that, if $\gamma = 0$, then $\mathcal{J} = 1$, and this case must be excluded for $F_q$ to be invertible (see section \ref{sec:networksasym}). Moreover, the same type of behaviour would have been observed if we had examined the limit $|\gamma| \rightarrow \infty$, for which $\mathcal{J}\rightarrow -1$. Hence, we only need to impose the existence of a unique absolute maximum for $0 < |\gamma| < \infty$. Crucially, this does not imply that the scheme with $\gamma = 0$ is useless. Figure \ref{priornetwork}.v shows that this scheme is giving information about the combination $\theta_2 + \theta_1 = \pi m$, with $m = 0, \pm 1, \pm 2, \dots$, that is, about the sum of the parameters. In fact, this can be readily seen by inserting $\gamma = 0$ in equation (\ref{multilikelihood}), since then the likelihood for a single shot is only sensitive to the equally weighted sum of the parameters. The calculations in the next section will reveal that while the asymptotic performance of this scheme is poor, it can be useful when $\mu$ is low.

\subsection{The role of inter-sensor correlations II}
\label{subsec:correlationsmultibayes}

Given the quantum strategy in equations (\ref{twonetworktransformed}) and (\ref{localpovm}) for a two-parameter qubit network, we wish to estimate two global properties of such network when the experiment operates both in and out of the regime of limited data. In particular, consider the linear functions $f_1(\boldsymbol{\theta}) = (2\theta_1 + \pi \theta_2)/\sqrt{4+\pi^2}$ and $f_2(\boldsymbol{\theta}) = (2\theta_1 + \theta_2)/\sqrt{5}$, which can be encoded in the columns of $V$ as
\begin{equation}
V = \frac{1}{\sqrt{20 + 5\pi^2}}
\begin{pmatrix}
2\sqrt{5} & 2\sqrt{4+\pi^2}\\
\pi \sqrt{5} & \sqrt{4+\pi^2}
\end{pmatrix}.
\label{funlimiteddata}
\end{equation}
We assume that both functions are equally relevant, so that $\mathcal{W} = \mathbb{I}/2$, and that our prior knowledge is represented by the prior probability $p(\theta_1, \theta_2) = 4/\pi^2$, when $(\theta_1, \theta_2) \in [0, \pi/2]\times[0, \pi/2]$, and zero otherwise. The area associated with this prior assignment is sufficiently small for the square error to be a suitable figure of merit in phase estimation \cite{jesus2018, friis2017}, and, thanks to our analysis in section \ref{subsec:multiprioranalysis}, we know that it will allow us to perform the estimation unambiguously when the asymptotically optimal strategies are employed, since $\Delta_0 = \pi^2/4 < \pi^2$.

Let us start by comparing a local strategy with an entangled scheme that is asymptotically optimal. The former assumes that the experiment is arranged such that $\gamma = 1$, $\mathcal{J} = 0$, while to find the properties of the latter we need to recall our results in section \ref{subsec:intersensorasymp} for the asymptotic role of inter-sensor correlations. Equation (\ref{optgeolinkanalytical}) indicates that, for $d = 2$, 
\begin{equation}
\mathcal{J}_{\mathrm{opt}} = \left(1- \sqrt{1 - \mathcal{G}^2}\right)/\mathcal{G},
\label{twoparametergeolink}
\end{equation}
when $\mathcal{G}\neq 0$, and $\mathcal{J}_{\mathrm{opt}} = 0$ if $\mathcal{G} = 0$. In addition, $\mathcal{J} = (1-\gamma^2)/(1+\gamma^2)$, and by combining the latter expression with equation (\ref{twoparametergeolink}) we find that
\begin{equation}
\gamma_{\mathrm{opt}} = \pm \left(\frac{\mathcal{G}-1+\sqrt{1-\mathcal{G}^2}}{\mathcal{G} + 1 - \sqrt{1-\mathcal{G}^2}}\right)^{\frac{1}{2}},
\label{gammaoptlink}
\end{equation}
when $\mathcal{G}\neq 0$, and $\mathcal{\gamma}_{\mathrm{opt}} = 1$ if $\mathcal{G} = 0$. The normalisation term for the functions in equation (\ref{funlimiteddata}) is simply $\mathcal{N} = \mathrm{Tr}(\mathcal{W} V^\transpose V) = 1$, while the geometry parameter is 
\begin{equation}
\mathcal{G} = \frac{1}{\mathcal{N}} \mathrm{Tr}\left(\mathcal{W} V^\transpose \mathcal{X} V \right) = \frac{8 + 10\pi + 2\pi^2}{20 + 5\pi^2} \approx 0.853.
\end{equation}
By inserting this result in equations (\ref{twoparametergeolink}) and (\ref{gammaoptlink}) we have that $\gamma_\mathrm{opt} \approx \pm 0.531$ (we can choose the positive solution without loss of generality) and $\mathcal{J} = 0.561$, where the latter verifies that this state is indeed entangled (note that the two-sensor state in equation (\ref{gammastatetwo}) is only separable when $\gamma^2 = 1$).   

\begin{figure}[t]
\centering
\includegraphics[trim={0.2cm 0cm 0.5cm 0cm},clip,width=14.75cm]{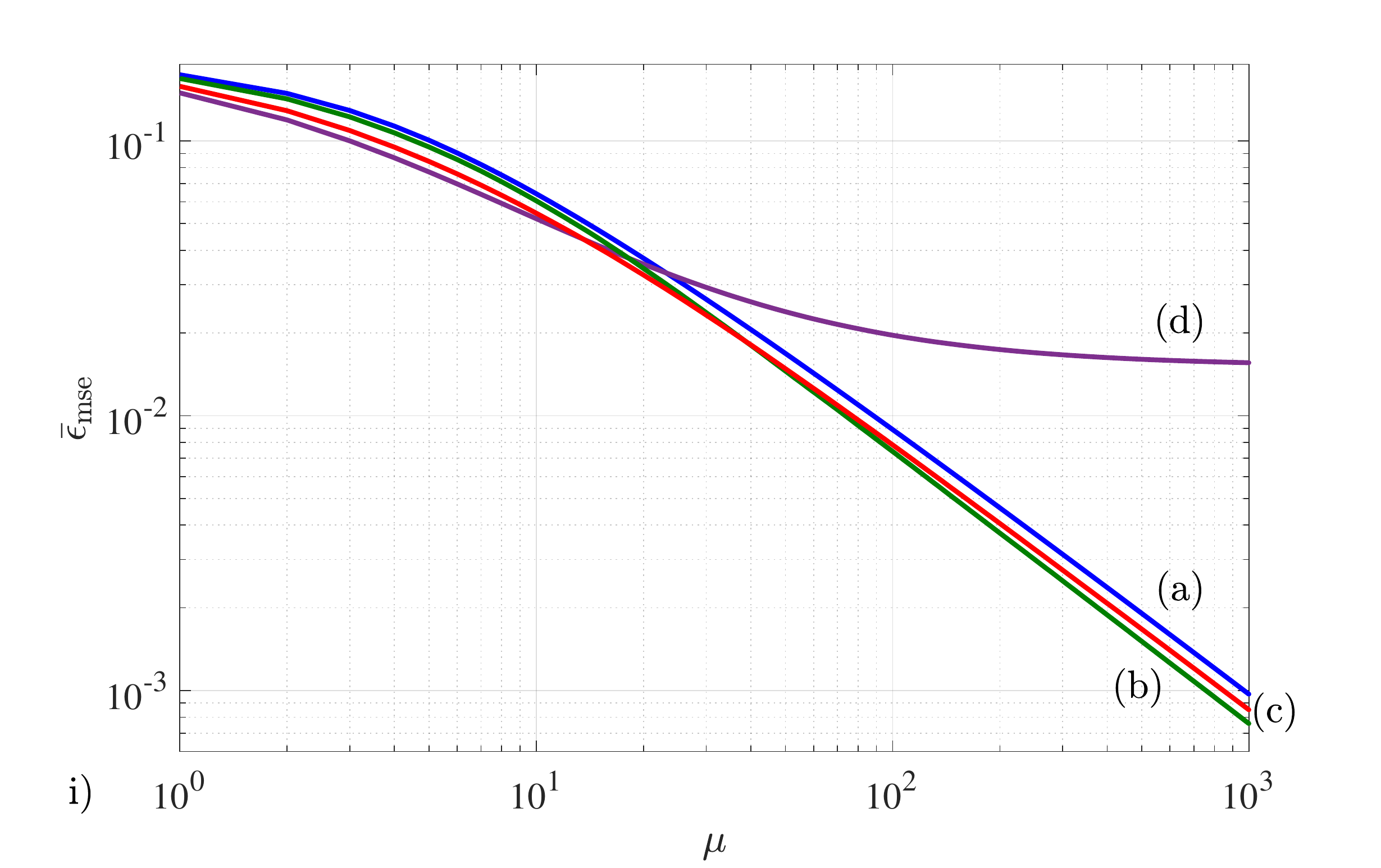}
\includegraphics[trim={0.2cm 0cm 0.5cm 0cm},clip,width=5.1cm]{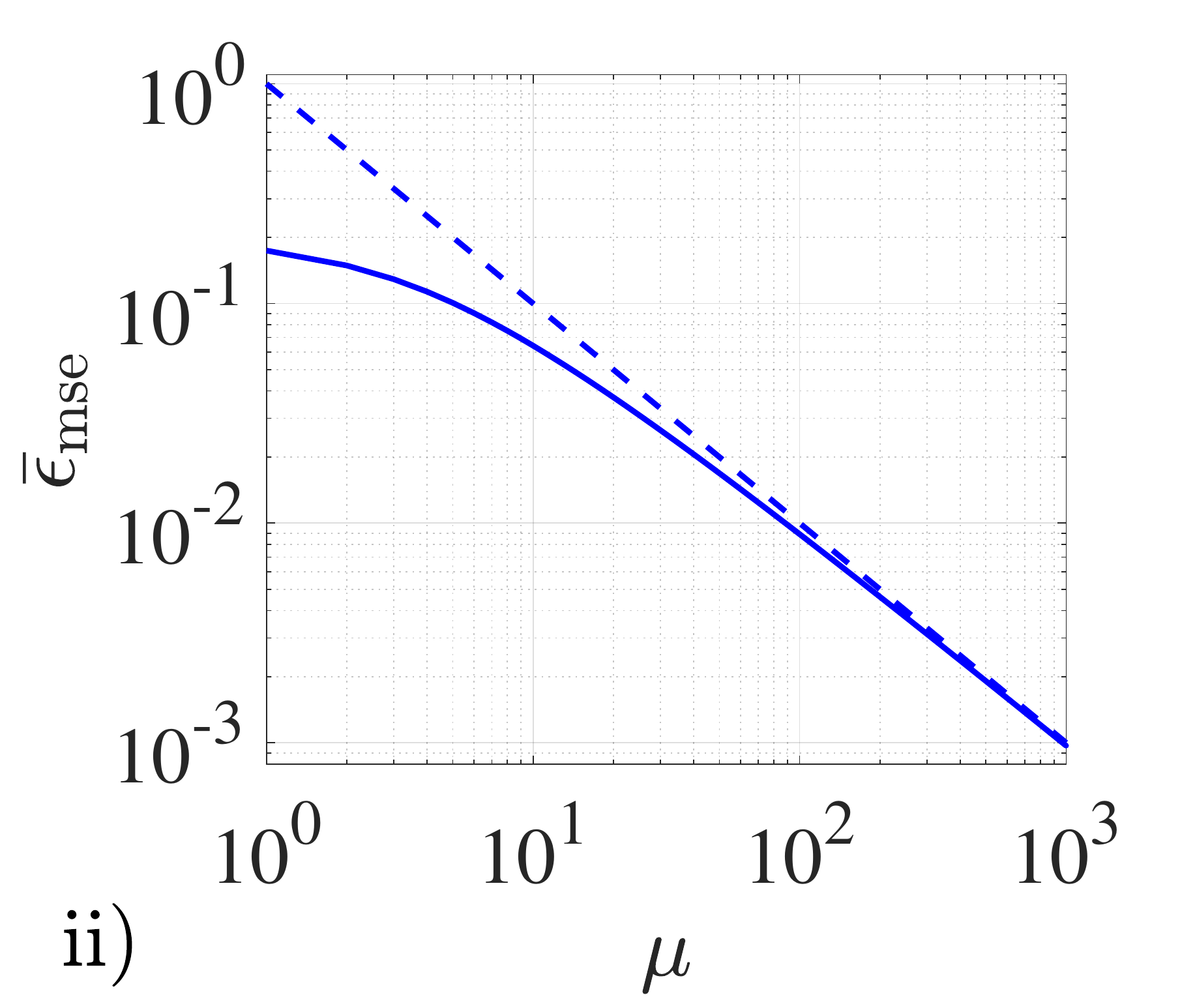}\includegraphics[trim={0.2cm 0cm 0.5cm 0cm},clip,width=5.1cm]{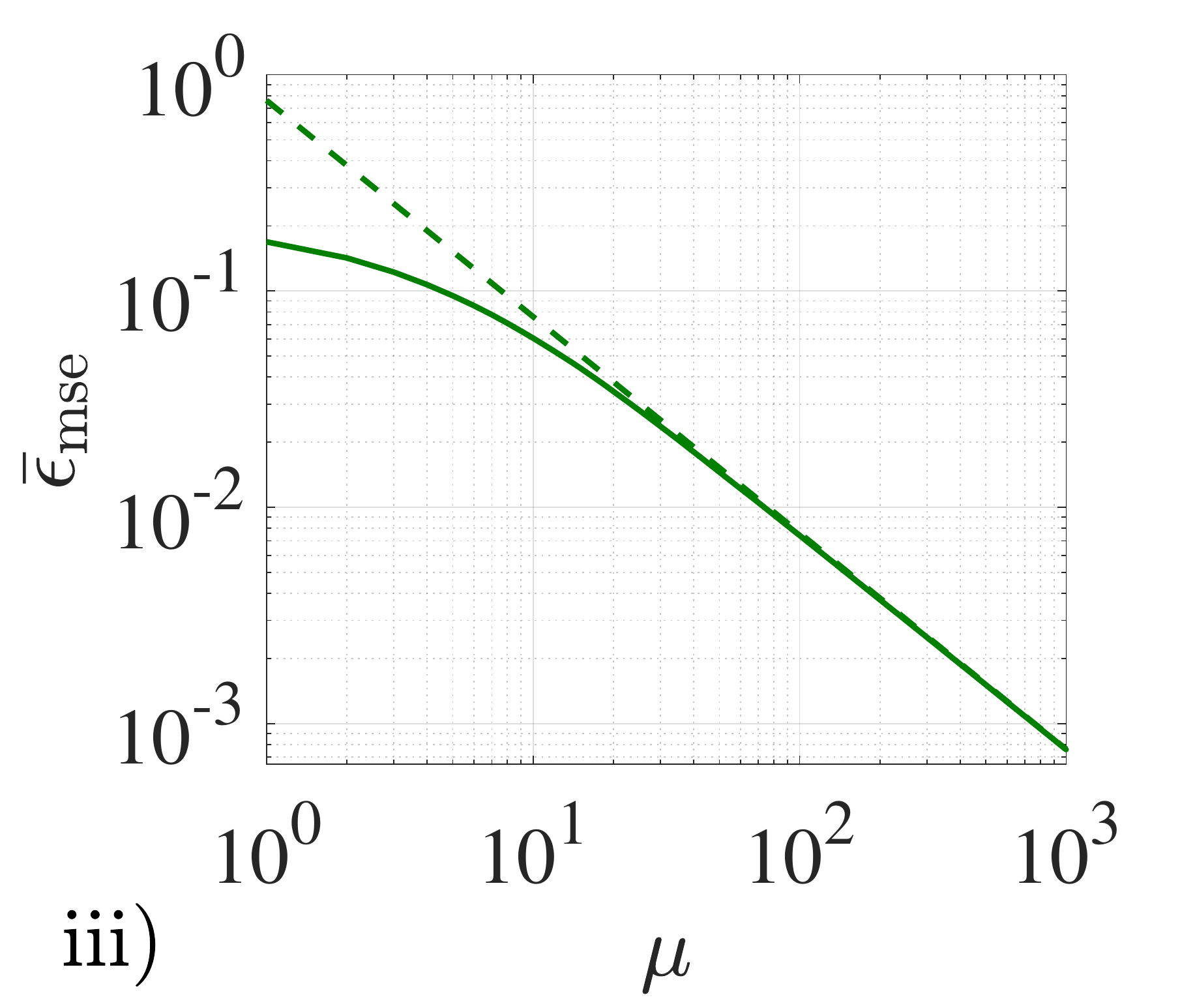}\includegraphics[trim={0.2cm 0cm 0.5cm 0cm},clip,width=5.1cm]{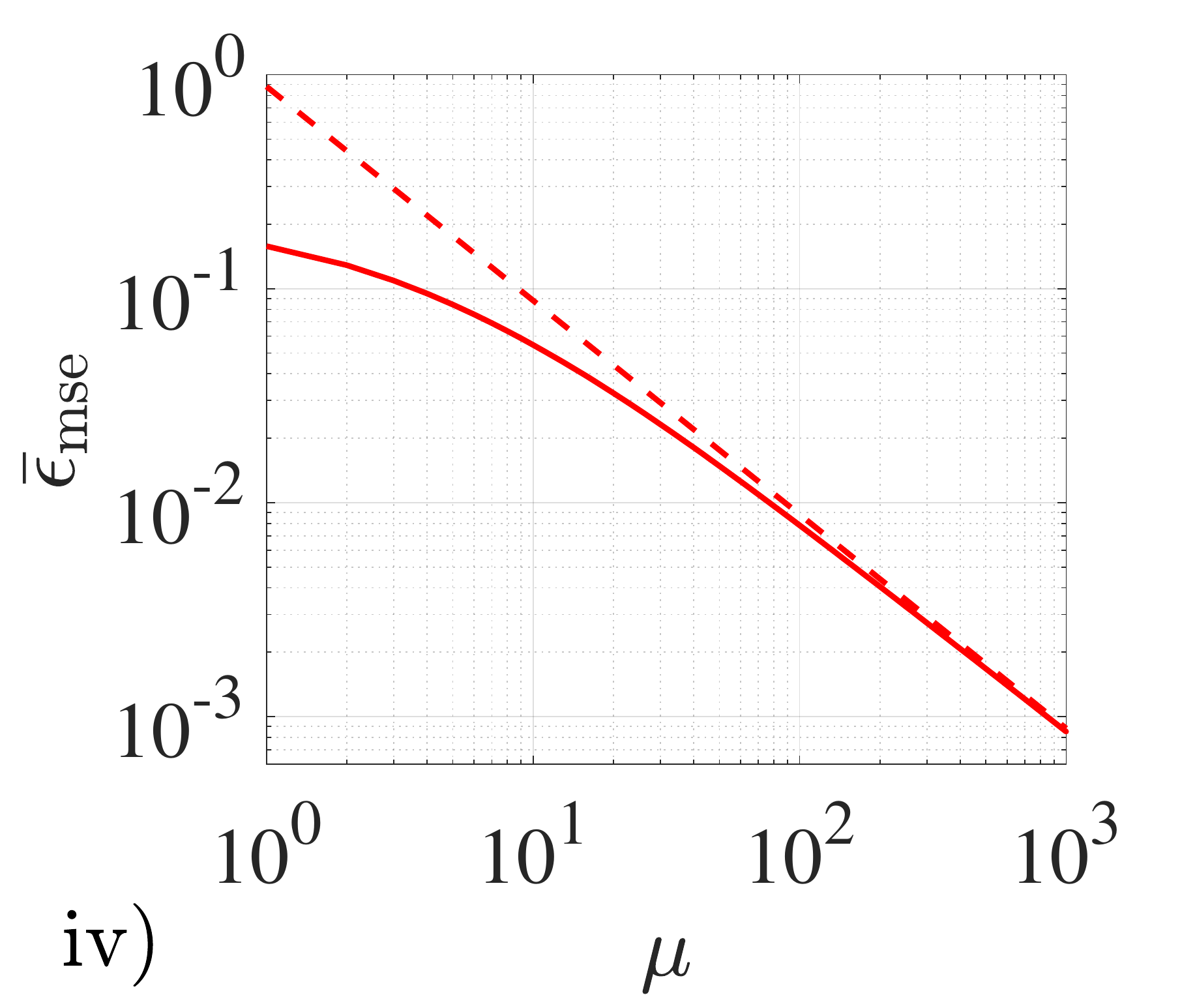}
\caption{i) Mean square error for the estimation of the linear functions $f_1(\boldsymbol{\theta}) = (2\theta_1 + \pi \theta_2)/\sqrt{4+\pi^2}$ and $f_2(\boldsymbol{\theta}) = (2\theta_1 + \theta_2)/\sqrt{5}$ by means of the two-sensor qubit network introduced in the main text, where (a, blue line) is a local strategy, with $\gamma = 1$, $\mathcal{J} = 0$; (b, green line) is the asymptotically optimal entangled strategy, with $\gamma = 0.531$, $\mathcal{J} = 0.561$; (c, red line) is a strategy whose enhancement has been balanced between the asymptotic and non-asymptotic regimes, with $\gamma = 0.334$, $\mathcal{J} = 0.799$; and (d, purple line) is a maximally entangled state, with $\gamma = 0$, $\mathcal{J} = 1$. Figures (ii - iv) compare the mean square error (solid lines) and the multi-parameter quantum Cram\'{e}r-Rao bound (dashed lines) for the strategies in (a - c), respectively, verifying that the latter is recovered asymptotically. All the calculations assume the weighting matrix $\mathcal{W} = \mathbb{I}/2$ and a flat prior of area $\Delta_0 = \pi^2/4$ centred around $(\pi/4, \pi/4)$.}
\label{nonasymptoticnetwork}
\end{figure}

Next we perform the numerical calculation of the Bayesian uncertainty $\bar{\epsilon}_{\mathrm{mse}}$ in equation (\ref{msefunctions}) for these two sensor-symmetric states, whose form as a function of $\gamma$ is in equation (\ref{gammastatetwo}); the measurement $E(n_i,k_i)=\ketbra{n_i,k_i}$ in equation (\ref{localpovm}) for the $i$-th repetition in a sequence of $\mu$ trials; and the optimal estimators
\begin{align}
\begin{pmatrix}
\tilde{f}_1(\boldsymbol{n}, \boldsymbol{k}) \\
\tilde{f}_2(\boldsymbol{n}, \boldsymbol{k}) 
\end{pmatrix} =\hspace{0.2em}&
\frac{4}{\pi^2\sqrt{20 + 5\pi^2}}
\begin{pmatrix}
2\sqrt{5} & \pi \sqrt{5}\\
2\sqrt{4+\pi^2} & \sqrt{4+\pi^2}
\end{pmatrix}
\\ \nonumber
&\times \int_0^{\pi/2} d\theta_1 \int_0^{\pi/2} d\theta_2 \hspace{0.2em}p(\boldsymbol{n}, \boldsymbol{k}|\theta_1, \theta_2)
\begin{pmatrix}
\theta_1 \\
\theta_2
\end{pmatrix},
\label{funlimiteddataest}
\end{align}
which arise from equation (\ref{fundoptest}) after inserting equation (\ref{funlimiteddata}). The results have been represented in figure \ref{nonasymptoticnetwork}.i as graphs (a) for the local scheme and (b) for the optimal entangled strategy. We can observe that the local strategy performs worse than the entangled one for any number of repetitions. Therefore, in this case we have that the prediction made by the asymptotic theory is qualitatively preserved in the non-asymptotic regime. However, a closer analysis reveals that the distance between the two lines is considerably less when $1 \leqslant \mu \lesssim 20$ than when $\mu \gg 1$, and this behaviour is reminiscent of that of a Mach-Zehnder interferometer \cite{jesus2018}. Indeed, optical probes with a large Fisher information (and thus a good asymptotic performance) have sometimes an error very close to that of a coherent laser beam in the regime of limited data, and coherent probes can be seen as an optical analogue of the notion of local strategy in this work. Moreover, the optical study in \cite{jesus2018} also demonstrated that a better asymptotic error is sometimes associated with a worse performance in the regime of low $\mu$. As a consequence, a natural question is whether a similar phenomenon can be exploited here, so that we can obtain an uncertainty that is lower than the error for the asymptotically optimal entangled state when the network operates in the non-asymptotic regime. 

To test this idea, let us select a third arrangement with an asymptotic error that lies between those of the local scheme and the asymptotically optimal strategy. The asymptotic error for our network can be written in terms of $\gamma$ as (see equations (\ref{symmetricfunctionssecond}) and (\ref{geometrylinkfactor}))
\begin{equation}
\bar{\epsilon}_{\mathrm{cr}} = \frac{\left(1+\gamma^2\right)\left[\left(1-\mathcal{G}\right)+\left(1+\mathcal{G}\right)\gamma^2\right]}{4\mu \gamma^2} \equiv \bar{\epsilon}_{\mathrm{qbit}}\left(\gamma\right).
\label{crmsetwoqubitnetwork}
\end{equation}
Using this we can find the value of $\gamma$ for the strategy satisfying our desideratum above by imposing that
\begin{equation}
\bar{\epsilon}_{\mathrm{qbit}}\left(\gamma\right) = \frac{1}{2}\left[\bar{\epsilon}_{\mathrm{qbit}}\left(\gamma_{\mathrm{loc}}=0\right) + \bar{\epsilon}_{\mathrm{qbit}}\left(\gamma_{\mathrm{ent}}=0.531\right)\right],
\end{equation}
and the solutions are $\gamma \approx \pm 0.334 , \pm 0.842$. So we take our third strategy to be the state in equation (\ref{gammastatetwo}) with $\gamma = 0.334$ (and thus $\mathcal{J} = 0.799$), a choice motivated by the fact that this is the option with the lower uncertainty for a single shot (in particular, $\bar{\epsilon}_{\mathrm{mse}}(\mu = 1, \gamma = 0.334) \approx 0.158$ and $\bar{\epsilon}_{\mathrm{mse}}(\mu = 1, \gamma = 0.842) \approx 0.173$). 

The uncertainty $\bar{\epsilon}_{\mathrm{mse}}$ for the third scheme has been represented as a function of the number of trials in figure \ref{nonasymptoticnetwork}.i, where it is labelled as (c). As expected, this error lies equidistantly between the local and the asymptotically optimal strategies when $\mu \gg 1$, but this is no longer the case in the regime of limited data. More concretely, the graphs for the asymptotically optimal strategy and the new scheme cross each other when $\mu \approx 40$, so that the former is optimal when $\mu > 40$ and the latter is the preferred choice if $1 \leqslant \mu \lesssim 40$. Consequently, we may say that trading a part of the asymptotic enhancement is sometimes associated with an improved performance in the non-asymptotic regime, which constitutes a multi-parameter generalisation of the analogous phenomenon in \cite{jesus2018} for a Mach-Zehnder interferometer.

\begin{table} [t]
\centering
{\renewcommand{\arraystretch}{1.2}
\begin{tabular}{|l|c|c|c|}
\hline
Strategy & $\gamma$ & $\mathcal{J}$ & $\mu_{\tau} (\Delta_0=\pi^2/4)$ \\
\hline
\hline
Local & $1$ & $0$ & $4.58\cdot 10^2$ \\
Asymptotically optimal & $0.531$ &$0.561$ & $4.3\cdot 10$ \\
Balanced enhancement & $0.334$ & $0.799$ & $5.37\cdot 10^2$  \\
Maximally entangled & $0$ & $1$ & $-$ \\
\hline
\end{tabular}}
\caption{Properties of different strategies based on a two-parameter qubit network, where $\gamma$ selects the state and $\mathcal{J}$ is the amount of inter-sensor correlations. The POVM is separable for all four schemes, but only the local strategy is based on a separable state. The asymptotically optimal strategy minimises the quantum Cram\'{e}r-Rao bound. The balanced strategy has also been enhanced via quantum correlations, but it is not asymptotically optimal because part of this enhancement has been traded to instead enhance its non-asymptotic performance. Finally, the fourth strategy uses a maximally entangled state. We note that the fourth column provides the number of repetitions $\mu_{\tau}$ needed such that the relative error between the Bayesian uncertainty and the Cram\'{e}r-Rao bound is equal to or less than a $5\%$ threshold (see \cite{jesus2017}), and in general it depends on the available prior information. Importantly, this calculation does not apply to the strategy with a maximally entangled state, since the estimation uncertainty for the latter does not have an asymptotic limit in the sense of section \ref{hybrid}. These results demonstrate the state-dependent nature of the conditions required to approach the Cram\'{e}r-Rao in multi-parameter systems.}
\label{multinetworkstable}
\end{table}

Interestingly, the balanced strategy ($\gamma = 0.334$, $\mathcal{J} = 0.799$), which provides a better precision in the non-asymptotic regime, is associated with larger inter-sensor correlations, and in what follows we propose a potential explanation for this. Let us first recall that, when $\mu$ is large, the information about the global properties is essentially provided by the measurement outcomes that we accumulate as $\mu$ grows, which contrasts with the non-asymptotic regime where the information is a mixture of prior knowledge and experimental data. This implies that the optimal correlation strength predicted by the asymptotic theory is implicitly assuming a large amount of information, while the information available in the non-asymptotic regime is poorer because $\mu$ is low and the prior in equation (\ref{multiprior}) is only moderately informative. It is thus reasonable to expect that the asymptotically optimal amount of entanglement \emph{is generally inappropriate}  in the non-asymptotic regime. One can then try to compensate the low amount of information in the regime with limited data by choosing $\mathcal{J}$ judiciously. In our case, we observe that our functions are clustered around the equally weighted sum of the parameters, since the geometry parameter of the former is $\mathcal{G} \approx 0.853$ and this is relatively close to the geometry parameter of the latter, $\mathcal{G} = 1$. In turn, this motivates choosing a $\mathcal{J}$ that is closer to that associated with $\boldsymbol{1}$, which is $\mathcal{J} = 1$, in order to enhance the precision when $\mu$ is low, and this is what (b) and (c) in figure \ref{nonasymptoticnetwork}.i show.

We may push this intuition further and consider a network with $\gamma = 0$, $\mathcal{J} = 1$, which makes the state in equation (\ref{gammastatetwo}) maximally entangled. Its graph has been labelled as (d) in figure \ref{nonasymptoticnetwork}.i, and upon comparing it with the three previous strategies we see that the maximally entangled state is the best option when $1 \leqslant \mu \lesssim 10$. The price that we pay for this low-$\mu$ enhancement is that the scheme ceases to be useful after $\mu \approx 20$ trials, and it is asymptotically beaten by the rest of schemes, including the local strategy. We notice that this result is consistent with our analysis in section \ref{subsec:multiprioranalysis}, where we established that this probe is only sensitive to the equally weighted sum of the parameters. 

The maximally entangled state also illustrates how, despite the lack of an asymptotic approximation in the sense of section \ref{hybrid}, we can still perform a Bayesian estimation using such strategy, even when it has limited usefulness. On the contrary, for the local, asymptotically optimal and balanced strategies we have that the Bayesian mean square errors converge to their respective Cram\'{e}r-Rao bounds, as it may be verified by observing figures \ref{nonasymptoticnetwork}.ii - \ref{nonasymptoticnetwork}.iv. The number of repetitions required for the relative error between these Bayesian uncertainties and the asymptotic bounds to be equal to or less than $5\%$ runs from $\mu \sim 10$ to $\mu \sim 10^2$ (see table \ref{multinetworkstable} for more details). 

In summary, in this section we have demonstrated that the strength of the inter-sensor correlations that is useful to estimate a given collection of global properties changes substantially for different amounts of data, i.e., for different values of $\mu$. Since this is the same type of behaviour that we had established for single-parameter schemes in \cite{jesus2018}, we \emph{conjecture} that the novel effects associated with a limited number of trials, which here have been uncovered using specific examples, are a general feature of non-asymptotic quantum metrology, and that they are generally present in a wide range of experiments operating in the regime of limited data.

\section{Summary and outlook}
\label{conclusions}

The central question addressed in this work has been that of the role of inter-sensor correlations in the estimation of linear functions with arbitrary geometry, having exploited a sensor-symmetric qubit network in the presence of different amounts of data. First we focused on the asymptotic part of the problem, and by optimising the class of sensor-symmetric states, we have established an optimal link between correlation strength and the geometry of the linear functions. Thanks to this we have been able to demonstrate that, while entanglement is useful for many geometrical configurations, it is sometimes detrimental even with functions that are non-trivial global properties. Furthermore, we have found that forms of entanglement other than those of a pairwise nature are in fact irrelevant in this regime. Hence, our approach significantly extends previous studies in networked quantum sensing that had only considered the estimation of a single function or a collection of $l = d$ orthogonal ones. 

Given that, in practice, the number of trials $\mu$ is always finite and possibly small, we have also performed a non-asymptotic analysis of sensing networks. To this end we have introduced a \emph{hybrid} estimation technique combining asymptotic and non-asymptotic optimisations in Bayesian estimation. This approximate but powerful approach has revealed that the correlation strength that is optimal for sensor-symmetric networks crucially depends on the number of times that we repeat the experiment. Additionally, we have demonstrated how the non-asymptotic precision may be enhanced by trading precision enhancements associated with the asymptotic regime. 

Admittedly, while many of our asymptotic results are valid for $d$ parameters, our Bayesian analysis has been restricted to the $d=2$ case due to numerical complexity. Hence, developing methods to overcome this limitation may have a major impact in the long run. For instance, it would be interesting to examine whether the irrelevancy of forms of entanglement other than those that generate pairwise correlations is also true for a low number of trials, which is a question that requires simulations where $d \geqslant 3$. One possibility is to modify the multi-parameter algorithm in \cite{jesus2019thesis} that we have exploited in section \ref{bayesianresults}, such that the integrals associated with the parameters $\boldsymbol{\theta}$ are performed with Monte Carlo techniques. Alternatively, we could employ some other quantum bound whose calculation is simple enough to study cases where both $\mu$ and $d$ are unrestricted. One potential candidate fulfilling the latter is the multi-parameter quantum Ziv-Zakai bound in \cite{zhang2014}, although, according to our findings in \cite{jesus2019b}, we cannot expect the results derived using this type of tool to be tight in general. 

Another important direction for future work is to extend our analysis to include the potential effect that decoherence may have in our conclusions. For example, it would be desirable to establish whether, in such case, inter-sensor correlations are still generally detrimental for the estimation of linear functions whose geometry parameter vanishes, i.e., when $\mathcal{G} = 0$. Note that our hybrid estimation technique can still be employed here, but replacing the ideal quantum Cram\'{e}r-Rao bound by its version for mixed states when such bound is applied to equation (\ref{ccrb}). 

Notwithstanding these limitations, our methodology has revealed new important aspects of the role of entanglement in the simultaneous estimation of linear functions with networked schemes, and these results could contribute decisively towards a powerful theoretical framework for networked quantum sensing.

\ack

JR performed and interpreted the calculations of his contribution to this work during his time at the University of Sussex. We thank Jasminder Sidhu, Matthew Thornton and George Knee for helpful discussions. JR and JAD acknowledge support from the South East Physics Network (SEPnet) and the United Kingdom EPSRC through the Quantum Technology Hub: Networked Quantum Information Technology (grant reference EP/M013243/1), and JR also from Engineering and Physical Sciences Research Council (EPSRC) grant EP/T002875/1. PAK acknowledges support from the Royal Commission for the Exhibition of 1851. All statements of fact, subjective views, opinions, or conclusions expressed herein are strictly those of the authors; they do not represent the official views or policies of the Department of Energy or the U.S. Government. Sandia National Laboratories is a multimission laboratory managed and operated by National Technology \& Engineering Solutions of Sandia, LLC, a wholly owned subsidiary of Honeywell International Inc., for the U.S. Department of Energy's National Nuclear Security Administration under contract DE-NA0003525.

\appendix

\section{Constructing the multi-parameter prior probability}\label{multipriorappsec}

Suppose that, according to our prior information about the network, we know that: a) a priori there is no reason to expect that the parameters $\boldsymbol{\theta}$ are correlated in any way with each other and b) we are ignorant of the magnitudes of the parameters, although c) only within a hypervolume $\Delta_0$ that is centred around $\boldsymbol{\bar{\theta}}=(\bar{\theta}_1, \dots, \bar{\theta}_d)$. The purpose of this appendix is to construct a prior density that codifies this state of information. 

Given a), the parameters are initially thought of as independent in the statistical sense, which in turn allows us to formalise b) as the assertion that a displacement by an arbitrary real vector $\boldsymbol{c}$ does not change our state of information. That is, $\boldsymbol{\theta}$ and $\boldsymbol{\theta}' = \boldsymbol{\theta} + \boldsymbol{c}$ generate equivalent estimation problems. 

At the same time, this invariance in our state of information is equivalent to imposing that $p(\boldsymbol{\theta})d\boldsymbol{\theta} = p(\boldsymbol{\theta}')d\boldsymbol{\theta}'=p(\boldsymbol{\theta}+\boldsymbol{c})d\boldsymbol{\theta}$, which gives rise to the functional equation $p(\boldsymbol{\theta}) = p(\boldsymbol{\theta}+\boldsymbol{c})$, and the latter can be satisfied with $p(\boldsymbol{\theta}) \propto 1$. 

Finally, c) indicates that the argument in the previous two paragraphs can only be approximately fulfilled in a portion of the parameter domain with hypervolume $\Delta_0$ centred around $\boldsymbol{\bar{\theta}}=(\bar{\theta}_1, \dots, \bar{\theta}_d)$. Since a priori the parameters are thought of as independent, we may express the hypervolume $\Delta_0$ as $\Delta_0 = \prod_{i=1}^d W_{0,i}$, where $W_{0,i}$ is the prior width for the $i$-th parameter. Therefore, our multi-parameter prior will be
\begin{equation}
p(\boldsymbol{\theta})=1/\Delta_0 = 1/\left(\prod_{i=1}^d W_{0,i}\right), 
\label{multipriorapp}
\end{equation}
for $\boldsymbol{\theta}\in [\bar{\theta}_1 - W_{0, 1}/2, \bar{\theta}_1 + W_{0, 1}/2]\times \cdots \times [\bar{\theta}_d - W_{0, d}/2, \bar{\theta}_d + W_{0,d}/2]$, and zero otherwise, which is the prior introduced in section \ref{ournetwork} and employed in the main text. We notice that this is a multi-parameter application of a method proposed by Jaynes to construct objective prior probabilities \cite{jaynes1968, jaynes2003}. Other methods can be found in \cite{kass1996, toussaint2011}.

\section{Optimising the multi-parameter Bayesian uncertainty: review of techniques}\label{multioptimisation}

There are several ways of addressing the problem of optimising the uncertainty in equation (\ref{msefunctions}) with respect to the estimators $\boldsymbol{\tilde{\theta}}(\boldsymbol{m})$, the measurement scheme $E(m_i)$ and the initial sensor-symmetric state $\rho_0$. One option is to perform a direct minimisation \cite{helstrom1976, helstrom1974, holevo1973b, holevo1973}, which is sometimes possible in covariant estimation \cite{chiara2003, chiribella2005, holevo2011, rafal2020} but generally intractable. Alternatively, one can bound the estimation error and search for the strategy that better approaches that bound, which may be attempted with tools such as the Yuen-Lax bound \cite{yuen1973}, the quantum Weiss-Weinstein bound \cite{tsang2016}, or some multi-parameter version of the quantum Ziv-Zakai bound \cite{zhang2014, berry2015}, among others \cite{gill2011, rafal2020}. This method usually suffers from the lack of tightness of the bounds, although this can be partially overcome with the Bayesian analogue of the Helstrom Cram\'{e}r-Rao bound that we recently constructed in \cite{jesus2019b} (see also \cite{jasminder2019, rafal2020}), since it can be saturated in certain cases and we showed how to exploit it for the estimation of local parameters (i.e., $\boldsymbol{\theta}$). Nevertheless, we have followed the weaker but computationally simpler hybrid approach in section \ref{hybrid} because the theory of estimating global properties of a network is more challenging, and we leave the application of more sophisticated methods to the estimation of linear functions for future work. 

\section{Minimisation of the asymptotic uncertainty for linear functions}
\label{asymlinearmin}

The optimal strength for the inter-sensor correlations in equation (\ref{optgeolinkanalytical}) can be found as follows. If we look at $\bar{\epsilon}_{\mathrm{f}} = \mathcal{N}h\left(\mathcal{J}, \mathcal{G}, d\right)/\mu$ as a function of $\mathcal{J}$, where we recall that, according to the discussion in section \ref{sec:networksasym},
\begin{equation}
h\left(\mathcal{J}, \mathcal{G}, d\right) = \frac{\left[1 + (d-2 - \mathcal{G})\mathcal{J}\right]}{(1 -\mathcal{J})[1+(d-1)\mathcal{J}]},
\label{geometrylinkfactorapp}
\end{equation}
then the equation for its extrema is
\begin{equation}
\frac{\mathcal{N}}{\mu}\frac{\partial h\left(\mathcal{J}, \mathcal{G}, d\right)}{\partial \mathcal{J}} = \frac{\mathcal{N}}{\mu} \frac{(d-1)(d-2-\mathcal{G})\mathcal{J}^2+ 2(d-1)\mathcal{J} - \mathcal{G}}{(1 -\mathcal{J})^2[1+(d-1)\mathcal{J}]^2} = 0,
\label{slope}
\end{equation}
whose solutions are
\begin{equation}
\mathcal{J}_{\pm} = \frac{1}{\mathcal{G}+2-d}\left[1\mp \sqrt{\frac{(\mathcal{G}+1)(d-1-\mathcal{G})}{d-1}}\right].
\end{equation}
Since we need to restrict our study to the range $1/(1-d)<\mathcal{J} < 1$ for $F_q$ to be invertible, only $\mathcal{J_{+}}$ is a valid candidate to find a minimum. Next we examine the sign of the slope in the left hand side of equation (\ref{slope}) for some values of $\mathcal{J}$ around $\mathcal{J_{+}}$. By noticing that $\mathcal{N}/\mu > 0$ and using the endpoints of the domain for $\mathcal{J}$ we find that
\begin{equation}
\frac{\partial h\left(1 - \varepsilon, \mathcal{G}, d\right)}{\partial \mathcal{J}} > 0, ~~\frac{\partial h\left(1/(1-d) + \varepsilon, \mathcal{G}, d\right)}{\partial \mathcal{J}} < 0
\end{equation}
for an arbitrarily small $\varepsilon > 0$ when $\mathcal{G}\neq -1$, $\mathcal{G}\neq d-1$, which we exclude to guarantee that $\mathcal{J}\neq 1/(1-d)$, $\mathcal{J}\neq 1$. Consequently, $\mathcal{J_{+}}$ gives rise to the minimum that we were looking for.

\section{Eigendecomposition of $\mathcal{X}$}
\label{eigencalx}

The characteristic equation for $\mathcal{X} = \mathcal{I} - \mathbb{I}$ is 
\begin{equation}
\mathrm{det}\left(\mathcal{X} - \lambda\mathbb{I}\right) = \mathrm{det}\left[\boldsymbol{1}\boldsymbol{1}^\transpose - (1 + \lambda)\mathbb{I}\right] \propto \left(1 - d + \lambda\right)\left(1+\lambda\right)^{d-1} = 0, 
\end{equation}
giving the eigenvalues $\lambda_1 = d-1$, with multiplicity $1$, and $\lambda_2 = -1$, with multiplicity $d-1$ (see the calculations associated with equation (\ref{characteristicfim}), whose eigenvalues are obtained in the same way). By inspection we see that $\boldsymbol{1}$ is one of the eigenvectors. Since the latter satisfies that $\mathcal{X}\boldsymbol{1} = (\boldsymbol{1}\boldsymbol{1}^\transpose - \mathbb{I})\boldsymbol{1} = (d-1)\boldsymbol{1}$, the rest of the eigenvalues must be associated with the subspace orthogonal to $\boldsymbol{1}$, and this concludes the eigendecomposition of $\mathcal{X}$.


\section*{References}
\bibliographystyle{unsrt}

\bibliography{ref13052020}

\end{document}